%% file: document.tex
\documentclass[sn-mathphys-num]{sn-jnl}
\usepackage{graphicx}%
\usepackage{multirow}
\usepackage{amsmath,amssymb,amsfonts}
\usepackage{amsthm}
\usepackage{mathrsfs}
\usepackage[utf8]{inputenc}
\usepackage[title]{appendix}
\usepackage{xcolor}
\usepackage{textcomp}
\usepackage{manyfoot}
\usepackage{booktabs}
\usepackage{makecell}
\usepackage{algorithm}
\usepackage{algorithmicx}
\usepackage{algpseudocode}
\usepackage{listings}
\usepackage[most]{tcolorbox}
\usepackage{colortbl}
\usepackage[skip=10pt]{caption}
\usepackage{subcaption}
\PassOptionsToPackage{hyphenbreaks}{breakurl}

\lstdefinelanguage{xquery}{
	morekeywords={for,in,doc,where,not,some,satisfies, return},
	columns=fullflexible,
	breaklines=true,
	postbreak=\mbox{\textcolor{red}{$\hookrightarrow$}\space},
	sensitive=true
}
\lstdefinelanguage{sparql}{
	morecomment=[l][\color{olive}]{\#},
	morekeywords={select,where,filter,not,exists},
	columns=fullflexible,
	breaklines=true,
	postbreak=\mbox{\textcolor{red}{$\hookrightarrow$}\space},
	sensitive=true
}
\lstdefinelanguage{cypher}{
	morekeywords={match,where,not,exists,return},
	columns=fullflexible,
	breaklines=true,
	postbreak=\mbox{\textcolor{red}{$\hookrightarrow$}\space},
	sensitive=true
}

\tcbset{
	mycode/.style={
		colback=white,
		colframe=gray!50,
		boxrule=0.2mm,
		arc=2mm,
		left=2mm,
		right=2mm,
		top=0mm,
		bottom=0mm,
		listing only,
		enhanced,
	}
}

\begin{document}
	\title{Domain-Specific Data Quality Analysis Using Technology-Independent Query Templates}

	\newcommand{\footnoteref}[1]{\textsuperscript{\ref{#1}}}
	\newcommand{\subrow}[1]{\hspace{0.8em}\textcolor{black!60}{#1}}

	\providecommand{\q}[1]{``#1''}

	\renewcommand*{\figureautorefname}{Fig.}
	\newcommand{\lstlistingautorefname}{Lst.}
	\renewcommand*{\sectionautorefname}{Sec.}
	\let\subsectionautorefname\sectionautorefname
	\let\subsubsectionautorefname\sectionautorefname
	
	\author*[1,2]{\fnm{Arno} \sur{Kesper}}
	\email{arno.kesper@uni-marburg.de}
	
	\author[2]{\fnm{Lukas Sebastian} \sur{Hofmann}}
	\email{lukas.hofmann@uni-marburg.de}
	
	\author[1]{\fnm{Markus} \sur{Matoni}}
	\email{markus.matoni@gwdg.de}
	
	\author[2]{\fnm{Gabriele} \sur{Taentzer}}
	\email{taentzer@mathematik.uni-marburg.de}
	
	\affil[1]{\orgname{Verbundszentrale der GBV}, \orgaddress{\city{G\"{o}ttingen}, \country{Germany}}}
	\affil[2]{\orgname{Philipps-Universit\"{a}t Marburg}, \orgaddress{\country{Germany}}}

	
	\abstract{\input{text/0_abstract}}
	
	\maketitle	
	
	\input{text/1_introduction}
	\input{text/2_motivation}

	\input{text/3_basics}
	\input{text/4_concept}
	\input{text/5_design}

	\input{text/6_evaluation}

	\input{text/7_related-work}
	\input{text/8_conclusion}
	\input{text/0_acknowledgement}

	\bibliography{bibliography}
\end{document}

%% file: text/0_abstract.tex
In an increasingly data-driven world, effectively working with data depends heavily on its quality.
Quality analysis is a central aspect of data quality management.
As data quality is typically domain- and context-specific, the definition of quality requirements is primarily the responsibility of domain experts.
However, domain experts often lack the query language expertise needed to implement quality analyses.
Therefore, the process of defining quality analyses results in a resource-intensive workflow that requires the involvement of technical experts, effectively excluding domain experts from independently managing data quality.

To address this challenge, we present the Quality Pattern Model framework (QPM), a model-driven approach to define templates for data quality analyses that are independent of specific database technologies and application domains.
QPM can eliminate the need for deep technical expertise and prevent the need for defining quality analyses several times for different database technologies.

We present a proof-of-concept implementation of this approach for three database technologies:
XML, RDF, and Neo4j.
We evaluate the expressiveness of our approach, its applicability in the cultural heritage domain, and its usability by domain experts.
For this purpose, we conducted a qualitative user study and empirically collected quality problems in a catalog.
Our findings suggest that QPM matches and even exceeds the expressiveness of common database query languages.
Furthermore, the results indicate that our tool enables domain experts to define template-based quality analyses independently, without requiring support of IT experts.

%% file: text/1_introduction.tex
\section{Introduction}
\label{sec_introduction}

In today’s data-driven world, working effectively with data depends heavily on its quality, understood as 'fitness for use' \cite{wang1996}.
As the amount of data continues to grow across all domains, ensuring data quality has become a fundamental challenge \cite{explodingdata, wang1996}.
Regular quality analyses are indispensable to maintaining high-quality data \cite{pipino2002}.
However, data quality can vary significantly, depending on the domain of interest and the context of use \cite{strong1997}.
Data quality requirements are criteria that characterize high-quality data \cite{wang1996, pipino2002}.
A data quality requirement can be operationalized through one or more constraints that specify measurable conditions for assessing compliance \cite{foundation, pipino2002}.

As data quality is inherently domain-specific \cite{strong1997}, domain experts with detailed knowledge of the data are primarily responsible for deriving constraints based on data quality requirements \cite{pham2010}.
{\em Although domain experts usually know the quality requirements for their domain-specific data, they are often unable to express them in a machine-readable form.}

In addition, domains such as cultural heritage comprise a variety of projects that utilize different database technologies.
Although the quality requirements are often similar across several projects, they frequently need to be specified again for each database technology, as there is no overarching approach for defining domain-specific data quality that is independent of specific database technologies.
Requirements are specific to each database technology, domain, and database schema.
Sometimes, they must even be tailored to individual databases.
This is tedious since the same quality requirements often have to be specified for several database technologies.
{\em In summary, there is no out-of-the-box solution for data quality assessment that can easily be adapted to any database technology or domain \cite{wang1996, pipino2002}.}

Common database technologies come with the following query languages:
For example, XQuery\footnote{\label{xquery}\url{https://www.w3.org/TR/xquery/} (2026-06-25)} is used for XML~\cite{xml},
SQL~\cite{sql} for relational databases (RDB), SPARQL\footnote{\label{sparql}\label{note_sparql}\url{https://www.w3.org/TR/sparql11-query/} (2026-06-25)} for RDF~\cite{rdf}, and
Cypher\footnote{\label{cypher}\url{https://neo4j.com/docs/cypher-manual} (2026-06-25)} for property graph databases, primarily for Neo4j\footnote{\label{neo4j}\url{https://neo4j.com/} (2026-06-25)}.
Thus, domain experts need to \emph{coordinate with data engineers} who have the technical knowledge necessary to define quality analyses \cite{collaborative}.
This leads to costly, time-consuming and error-prone workflows requiring intensive, precise communication about dynamic quality requirements between stakeholders.

To address these challenges, our goal is to close the skill gap between domain experts and data engineers by providing a systematic approach to defining domain-specific data quality analyses.
This approach is independent of specific database technologies so that it does not require in-depth technical expertise.
In addition, quality analyses need to be defined only once so that domain experts will be able to adapt them to different database technologies.

To achieve this goal, we developed a \emph{model-driven approach}, the Quality Pattern Model (QPM) framework, that defines domain-specific quality analyses \emph{independent of database technology}.
QPM builds on the observation that many data quality analyses follow reoccurring patterns, since similar quality dimensions and data validation mechanisms recur across domains and data models~\cite{fox1994, haug2021}.
Quality issues can be identified by searching for constraint violations.
The QPM approach starts with formulating abstract quality analyses as QPM templates.
Domain experts can use these QPM templates to specify concrete quality analyses, called QPM instances, for specific database technologies and database schemas.
These QPM instances are then automatically translated into specific database queries.
Data analysts can use these queries to perform quality analyses on specific databases.
We present the following new contributions:

\begin{enumerate}
\item We first examine the database technologies RDB, XML, RDF, and Neo4j.
	For each database technology we examine a query language that is specifically optimized for its structure and characteristics.
	\item We present QPM, a model-driven approach to define domain-specific data quality analysis independent of database technologies and data formats.
	The proof-of-concept implementation of QPM supports three database technologies, namely XML, RDF and Neo4j.
	This implementation includes an API and a web frontend to serve as a user interface for domain experts.
	\item We evaluate QPM in terms of its expressiveness, applicability and usability for domain experts.
	The implementation was applied to four concrete databases with different database schemas and three different database technologies using research data from the cultural heritage domain.
	{\em We found that this approach is widely applicable and more expressive than standard query languages.
	It is also easy for domain experts to use, thanks to its intuitive user interface.}
\end{enumerate}

In an earlier paper \cite{kesper2020}, we presented the initial concept of our approach and a proof-of-concept for QPM that only supported XML technology.
Since then, we have reworked the approach, extending its expressiveness, support for RDF and Neo4j data, and a UI for domain experts.

The structure of the paper is as follows:
We motivate our work and present a running example in \autoref{sec_motivation}.
In \autoref{sec_analysis}, we compare several database technologies and their query languages.
The framework QPM is described in \autoref{sec_approach}.
In \autoref{sec_tool-support}, we describe our tool support for three different database technologies, namely XML, RDF, and Neo4j.
In \autoref{sec_evaluation}, we evaluate the applicability, expressiveness, and usability of our tool support.
We discuss related work in \autoref{sec_related-work}.
Finally, we conclude our paper in \autoref{sec_conclusion}.

%% file: text/2_motivation.tex
\section{Quality Assessment of Cultural Heritage Data}
\label{sec_motivation}

Data quality is a multifaceted concept that is typically described by a number of \textit{quality dimensions}~\cite{qualitysurvey, qualitydimensions, wang1996, wand1996, strong1997, batini2009, cai2015}.
Data quality assessment can find specific quality problems in one or more data quality dimensions.
Quality assessment includes quality analysis, typically implemented as queries formulated in a language matching the database technology used.
For example, XQuery is used to query XML data.
The development of new database technologies and the increasing prevalence of databases within and between institutions has made it common to manage multiple databases with different technologies within an institution.
This requires domain experts to have expertise in different query languages.
However, these experts often lack the knowledge necessary to define queries independently.
Therefore, they depend on data experts to perform the necessary analyses in the query languages of the technologies used.
This dependency on others can lead to complex and inefficient workflows.
To avoid this, a technology-light approach is needed to support domain experts in defining quality analyses independently.

\subsection{Quality Problems in Cultural Heritage Data}
\label{sec_quality-problems}

As an example, we examine data quality analyses in the cultural heritage domain.
Data is collected from several different institutions, and later integrated for harvesting.
Prior to integration, the data must be analyzed for quality problems.
To learn about the quality problems in description data of cultural heritage objects, we conducted six qualitative interviews and a workshop with 19 domain experts who deal with the acquisition, modeling, management and usage of various types of cultural heritage data from large institutions, such as the German Documentation Center for Cultural Heritage (DDK)\footnote{\url{https://www.uni-marburg.de/de/fotomarburg} (2026-06-25)} and the German Digital Library (DDB)\footnote{\label{ddb}\url{https://www.deutsche-digitale-bibliothek.de/?lang=en} (2026-06-25)}.
In the context of cultural heritage, a database collects descriptive data about human-made objects, such as books, paintings, photos, and videos.

We compiled a comprehensive catalog of data quality problems by systematically documenting various aspects of each problem, such as its impact on data quality, possible causes, and identification possibilities \cite{comprehensive_problem_specifications}.
The catalog includes 73 intrinsic data quality problems; simple examples %
are {\em illegal values}, {\em invalid links}, {\em imprecise data} and {\em misplaced information}.
The resulting catalog matches the 
problems described in the literature \cite{rahm_data_2000, laranjeiro_survey_2015, furber2010using, oliveira_taxonomy_2005, oliveira_formal_2005, kim_taxonomy_2003}.
The catalog suggests that the majority of quality problems manifest recurring patterns.

\input{text/2b_running-example_table_further-examples_concrete}

\autoref{tab_furtherExamples-concrete} presents simple examples of quality constraints expressed in natural language.
Domain experts typically do not possess the technical skills required to define data quality analyses directly using query languages.
Therefore, an approach is needed that enables domain experts to define quality constraints based on natural language.
The table shows example constraints with recurring logical structures, where only a small amount of information (underlined in the examples) must be adapted for different scenarios.
This observation led to the idea of providing reusable patterns that can be instantiated to constraints as needed. 
To illustrate this idea, we first present a concrete example.
Then, we examine the query languages of common database technologies to identify recurring concepts and structures. 

\input{text/2b_running_example}

%% file: text/2b_running-example_table_further-examples_concrete.tex
\begin{table*}[htbp]
	\centering
	\caption{
		Below are some textual examples of specific quality constraints for a cultural heritage database that includes artists and paintings.
		Similar structural constraints also occur in different scenarios when only the underlined values are changed.
	}
	\label{tab_furtherExamples-concrete}
	\begin{tabular*}{\linewidth}{l}
		\toprule
		Each \underline{painting} must have an associated \underline{artist}. \\
		Each \underline{artist} must have a \underline{first} and a \underline{last name}. \\
		There should not be multiple \underline{persons} with the same \underline{name}, \underline{birthdate} and
		\\ \hspace*{2em} \underline{birthplace} as they are most likely duplicates. \\
		A \underline{painting} must have at least \underline{one} \underline{artist} associated. \\
		A \underline{person} has at most \underline{two} \underline{biological parents}. \\
		The \underline{death date} of a \underline{person} must be within \underline{120 years} of the \underline{birth date}. \\
		The \underline{height} of a \underline{painting} must be between \underline{0.01 meter} and \underline{100 meters}. \\
		A \underline{date} must conform to the format \underline{YYYY/MM/DD}. \\
		The relationship from \underline{paintings} to \underline{artists} must be consistent with the \\ 
		\hspace*{2em} relationship in the opposite direction. \\
		The \underline{identifier} for each \underline{artist} and \underline{painting} must be unique in the database. \\
		\bottomrule
	\end{tabular*}
\end{table*}

%% file: text/2b_running_example.tex
\subsection{Running Example}
\label{sec_example}

In the following, we will examine the first example for a quality constraint from \autoref{tab_furtherExamples-concrete} in more detail.
The following simple quality constraint is intended to prevent one specific case of incomplete information.

\begin{tcolorbox}[mycode]
\vspace{-0,2cm}
\begin{lstlisting}[label=lst_example,caption={Quality constraint example from cultural heritage}]
Every painting must have at least one associated artist.
\end{lstlisting}
\vspace{-0,2cm}
\end{tcolorbox}

Introducing parameters allows the constraint to be abstracted from the underlying data format:

\begin{tcolorbox}[mycode]
\vspace{-0,2cm}
\begin{lstlisting}
Every <type> must have at least one associated <type2>.
\end{lstlisting}
\vspace{-0,2cm}
\end{tcolorbox}

This constraint template can be instantiated for a whole group of completeness constraints.
The example constraint in \autoref{lst_example} is an instantiation with the types \q{paintings} and \q{artists}.
To find violations of this constraint in a dataset, the following search query is used.
It will be automatically deduced from the given constraint.

\begin{tcolorbox}[mycode]
\vspace{-0,2cm}
\begin{lstlisting}
Search for <type>, that has no associated <type2>.
\end{lstlisting}
\vspace{-0,2cm}
\end{tcolorbox}

The query analyzes data to identify particular elements that are missing mandatory information.
In this example, the query searches for an element representing a painting element that lacks artist information.
To identify this type of completeness violation, we can define a query that searches for \textit{all painting nodes in the database without a link to an artist node}.

Now, an approach is required to map such templates using natural language to different query languages.
As a prerequisite, we next compare the query languages of common database technologies.

%% file: text/3_basics.tex
\section{Database Technologies}
\label{sec_analysis}

Database technologies are (semi-)structured formats for storing and managing data \cite{databasetechnology}.
The interpretation of the data values is determined by the data fields present in a database and how they are linked.
How this structure is implemented depends on the structural paradigm of the database technology.
Each database technology has a different way of specifying relationships and querying specific values.
Thus, each database technology has its own query language that fits its specific concepts.
Hence, each query language has its own features and syntax.
To determine which details can be abstracted, it is important to compare the database structures and query languages to identify commonalities.
To our knowledge, no prior comparisons have addressed the fundamental structural design of different database technologies.
Instead, existing comparisons emphasize database operations and their performance.

\subsection{Comparison of Database Technologies}
\label{subsec_dbstructure_comparison}

This paper focuses on the following database technologies: relational databases, XML \cite{xml}, RDF \cite{rdf}, and Neo4j\footref{neo4j}.
In \autoref{fig_structure} we present a general overview over the structures of these technologies using examples.
Specifically, we examine how the structure is defined by comparing its components, identifying data containers, data values, and relationships.
Elements are interconnected with each other by relationships, and data values are assigned to the elements.
Typically, types are assigned to elements and relations to ensure data integrity and efficient querying.

\input{text/3_table_database-diagrams.tex}

\paragraph{Relational Databases}
\q{A \emph{relational database} (RDB) is a collection of data items organized in formally described tables from which data can be accessed or reorganized in many different ways}~\cite{jatana2012survey}.
MySQL and Oracle Database are two of the most popular database management systems for RDBs~\cite{jatana2012survey}.
Due to their tabular structure, RDBs require rigid schemas.
This results in high consistency and high performance for searches~\cite{jatana2012survey}.
However, if the data is incompatible with a rigid schema, namely, if the data is semi-structured or unstructured, then significant performance and scalability losses occur compared to non-relational databases~\cite{malik2020comparative}.
RDBs are based on tables, i.e., columns and rows.
A column specifies the values of a certain type of data.
Each row contains related \emph{values}.
Rows are often referred to as records or tuples, and columns are referred to as attributes or fields\footnote{\url{https://docs.oracle.com/javase/tutorial/jdbc/} (2025-12-18)}.
A record functions as a \emph{structure element}, and its fields contain the values.
Each record is identified by a key, which can be a unique ID.
A \emph{relationship} between records, possibly in different tables, is expressed through key references~\cite{kappel2001xml}.
The most commonly used query language for RDBs is the \emph{Structured Query Language} (SQL)~\cite{sql}.

\paragraph{XML}
The \emph{Extensible Markup Language} (XML)~\cite{xml} defines a data format structured in a hierarchical way.
XML is primarily used for data interchange and web services because it is ``self-describing'', human-readable, and more concise and flexible than RDBs~\cite{relationaltoxml}.
The \emph{structure} of an XML document forms a rooted tree structured by named elements called tags.
Tags can contain \emph{values} as attributes or as a leaf element.
XML supports two kinds of \emph{relationships}:
First, there are navigational relationships between elements via their relative location in the tree structure.
Second, XML provides the concept of key references (\emph{keyref}s) that use identifiers to point to specific nodes within the XML tree.
These references are similar to the relationships in a relational database that use foreign keys~\cite{kappel2001xml}.
Query languages for XML include XQuery, which specializes in data extraction, and XSLT\footnote{\url{https://www.w3.org/TR/xslt/} (2025-12-18)}, which specializes in data transformation.

\paragraph{RDF}
Graph databases are optimized for highly interconnected data.
They offer very flexible schemas and efficient handling of complex relationships, though they sacrifice scalability and performance~\cite{graphrelational}.
The Resource Description Framework (RDF) \cite{rdf} is a graph database technology based on triples in the \emph{subject-predicate-object} format.
Subjects refer to nodes, or \emph{structural elements}, that are identified by Internationalized Resource Identifiers (IRIs).
Objects can be either nodes or \emph{values}.
They are connected via predicates, which are referred to as \emph{relationships}.
These predicates connect nodes and values to a graph structure with values as leaf nodes.
The query language for RDF data is SPARQL (SPARQL Protocol and RDF Query Language).
SPARQL can query ``required and optional graph patterns along with their conjunctions and disjunctions''\footnoteref{note_sparql}.

\paragraph{Neo4j}
Another popular technology for graph databases is Neo4j\footref{neo4j}.
A Neo4j database contains nodes as basic \emph{structure elements} that are connected to each other via edges (specifying \emph{relationships}).
Nodes and edges can be assigned \emph{values} in the form of named properties.
A notable feature of this database technology is edge properties, which allow assigning values to relationships.
Neo4j databases and other property graph databases are queried using the Cypher language.
Although ISO is developing GQL \cite{gql} as a standardized, vendor-neutral alternative with similar syntax, limited adoption means Cypher remains the de facto standard in practice.

\paragraph{Comparison}
A feature-based comparison of the database technology structures is presented in \autoref{tab_technologycomparison}.
The following features were identified: structure, structure element, data value, and relationship.
Each database technology defines a structure for organizing data (values)~\cite{radoev2017comparison}.
This is accomplished using structure elements, such as columns in RDBs, tags in XML, IRIs in RDF, and nodes in Neo4j.
Data values are assigned to the structure elements as properties or attributes.
Neo4j also offers the option of assigning values to relationships using edge properties.
However, since our approach focuses on the commonalities between different technologies, we do not consider this feature further.

\input{text/3_table_structure-comparison.tex}

\subsection{Comparison of Query Languages}
\label{subsec_querycomparison}

Database technologies come with specific query languages appropriate for their supported data structures.
Query languages allow users to define queries for data retrieval with various features.
Data quality analysis is a specific form of data retrieval queries.

Looking at the running example from \autoref{sec_example}, we reconsidered the following constraint template:

\begin{tcolorbox}[mycode]
\vspace{-0,2cm}
\begin{lstlisting}
Every <type> must have at least one associated <type2>.
\end{lstlisting}
\vspace{-0,2cm}
\end{tcolorbox}

This constraint template can be instantiated for different database technologies, using different query languages.
For example, looking at a fictitious database file \textit{database.xml} in XML format, the constraint could be instantiated to the XQuery expression in \autoref{lst_example-xml}.

\begin{tcolorbox}[mycode]
\vspace{-0,2cm}
\begin{lstlisting}[language=xquery, label=lst_example-xml, caption={XML/XQuery query for the running example in \autoref{sec_example}}]
for $painting in doc(database.xml) // painting
where not { 
	some $artist in $painting / artist
	satisfies true()
}
return $painting
\end{lstlisting}
\vspace{-0,2cm}
\end{tcolorbox}

Considering Neo4j databases, it could be instantiated to the Cypher query language in \autoref{lst_example-neo}.
This constraint can also be expressed in other query languages, such as SPARQL (for RDF) and SQL (for RDB), depending on the database technology and format.

\begin{tcolorbox}[mycode]
\vspace{-0,2cm}
\begin{lstlisting}[language=cypher, label=lst_example-neo, caption={Neo4j/Cypher query for the running example in \autoref{sec_example}}]
match (painting:Painting)
where not ( exists {
	match (painting)-[:created_by]-(artist:Artist)
	})
return painting
\end{lstlisting}
\vspace{-0,2cm}
\end{tcolorbox}

In \autoref{tab_querylanguagecomparison}, we compare query features that are important for data quality analyses across the languages.

\input{text/3_table_query-comparison.tex}

The four considered query languages are all based on different algebras using different data traversal techniques, fitting to their supported data structures.
SQL (for relational databases) is based on relational algebra using tabular joins to access different parts of the data.
XQuery traverses XML data using XPath-expressions using tree algebra.
SPARQL defines its own algebra using property paths.
Lastly, Cypher is based on property graph algebra while relying on pattern matching for data traversal.
The expressiveness of all these different algebras strictly exceeds first-order logic.

Conditional data retrieval is based on defining strict conditions using operators.
Key operators for data quality analysis include the \emph{count} operator, e.g., for cardinality constraints, and \emph{comparison} operators for matching values against fixed thresholds.
To check the format (e.g., URLs, IRIs) of specific values, typically a \emph{regular expression} operator is used.
Each of these operator types is completely covered by all considered query languages.

Further operators are required for comprehensive quality analyses.
For example, values must be compared to external data, and the semantic understanding of the data must be checked using NLP techniques.
These capabilities exceed those of the query languages under consideration.

In summary, query languages have nearly equal \emph{expressive power} regarding data retrieval.

%% file: text/3_table_database-diagrams.tex
\begin{figure*}[thbp]
	\small
	\centering
	\begin{subfigure}[t]{.235\linewidth}
		\centering
		\includegraphics[width=.9\linewidth]{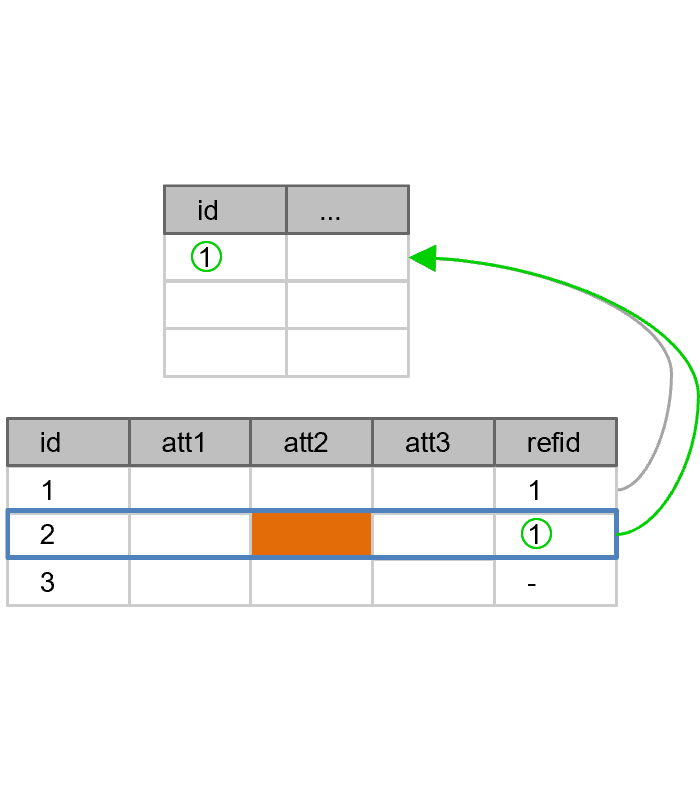}
		\caption{RDB}
	\end{subfigure}
	\begin{minipage}[t]{.01\linewidth}
	\end{minipage}
	\begin{subfigure}[t]{.235\linewidth}
		\centering
		\includegraphics[width=.9\linewidth]{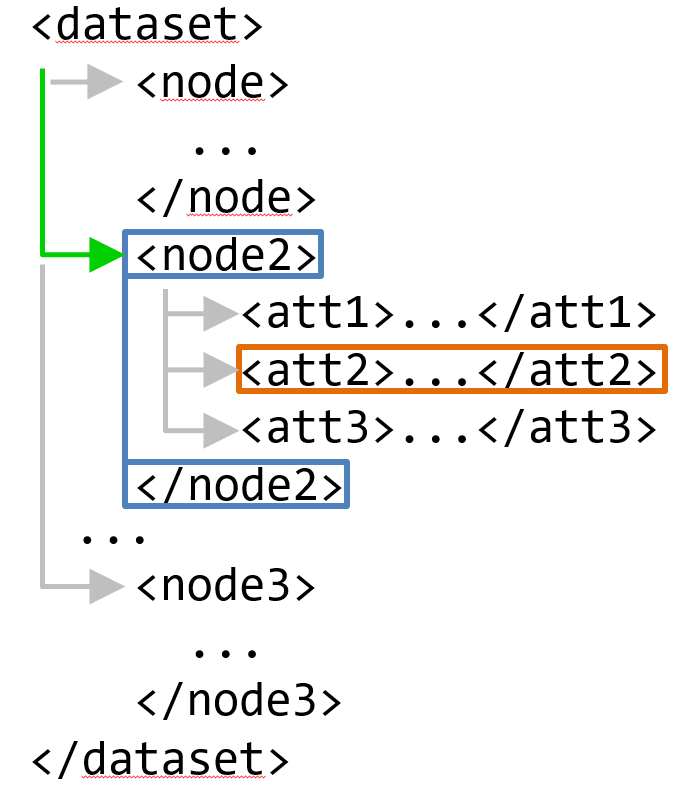}
		\caption{XML}
	\end{subfigure}
	\begin{minipage}[t]{.01\linewidth}
	\end{minipage}
	\begin{subfigure}[t]{.235\linewidth}
		\centering
		\includegraphics[width=.9\linewidth]{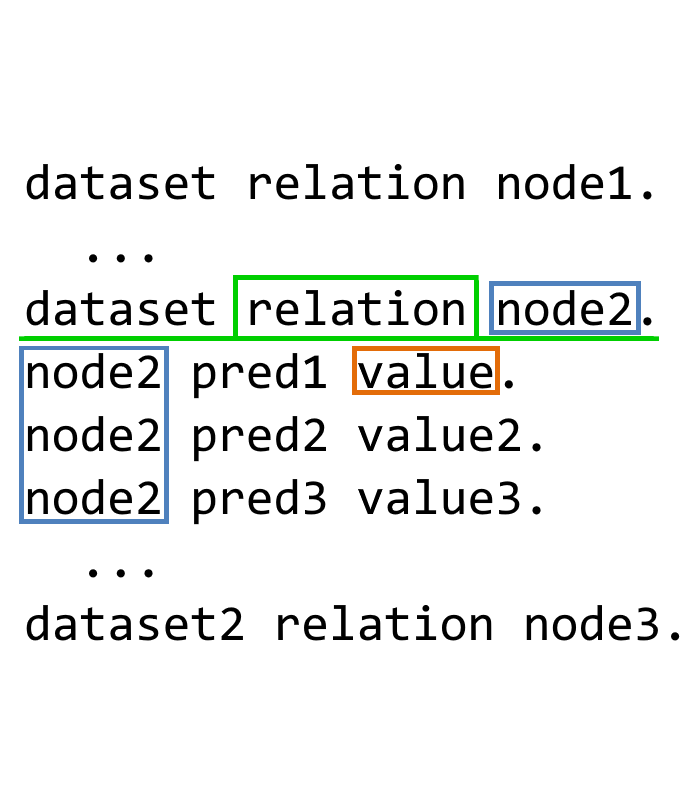}
		\caption{RDF}
	\end{subfigure}
	\begin{minipage}[t]{.01\linewidth}
	\end{minipage}
	\begin{subfigure}[t]{.235\linewidth}
		\centering
		\includegraphics[width=.9\linewidth]{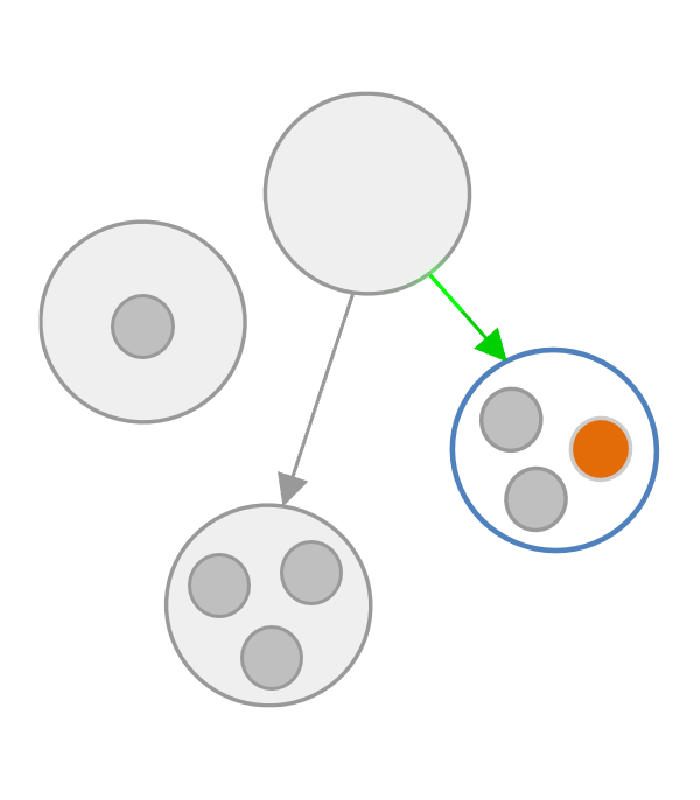}
		\caption{Neo4j}
	\end{subfigure}
	\vspace{-0.3cm}
	\caption{Comparison of database technology structures by example, \\ \normalfont		
		showing structure elements in blue, values in orange, and relations in green.
	}
	\label{fig_structure}
\end{figure*}

%% file: text/3_table_structure-comparison.tex
\begin{table*}[!h]
	\footnotesize
	\centering
	\caption{Feature-based comparison of database technology structures to highlight commonalities}
	\label{tab_technologycomparison}
	\begin{tabular}{@{}p{0.13\textwidth}p{0.17\textwidth}p{0.245\textwidth}p{0.155\textwidth}p{0.17\textwidth}@{}}
	\toprule
	                  & RDB        & XML          & RDF            & Neo4j          \\ \midrule
	Structure         & set of tables     & hierarchical & directed graph & directed graph \\
	Structure element & table row (tuple) & XML element  & IRI element   & node          \\
	Data value &
	  table entry &
	  value in tag or attribute &
	  value element &
	  property at a node or relationship \\
	Relationship &
	  tuple or \emph{keyref} reference &
	  structure navigation or \emph{keyref} reference &
	  predicate &
	  labeled relationship \\ \bottomrule
	\end{tabular}
\end{table*}

%% file: text/3_table_query-comparison.tex
\begin{table*}[htbp]
	\centering
	\caption{Feature-based comparison of different query languages regarding conditional data analysis}
	\label{tab_querylanguagecomparison}
	{
	\small
	\begin{tabular}{@{}lp{0.13\textwidth}p{0.175\textwidth}p{0.15\textwidth}p{0.15\textwidth}@{}}
		
	\toprule
	 &
	  SQL &
	  XQuery &
	  SPARQL &
	  Cypher \\ \midrule
	  \arrayrulecolor{gray!30}
	{FOL}-Exp. &
	  \multicolumn{1}{c}{\checkmark} &
	  \multicolumn{1}{c}{\checkmark} &
	  \multicolumn{1}{c}{\checkmark} &
	  \multicolumn{1}{c}{\checkmark} \\
	\subrow{Algebra} &
	  relational algebra & 
	  tree algebra & 
	  SPARQL algebra & 
	  property graph algebra \\
	\subrow{Traversal} &
	  joins &
	  XPath &
	  property paths &
	  pattern matching \\
	Count &
	  \multicolumn{1}{c}{\checkmark} &
	  \multicolumn{1}{c}{\checkmark} &
	  \multicolumn{1}{c}{\checkmark} &
	  \multicolumn{1}{c}{\checkmark} \\
	Comparison &
	  \multicolumn{1}{c}{\checkmark} &
	  \multicolumn{1}{c}{\checkmark} &
	  \multicolumn{1}{c}{\checkmark} &
	  \multicolumn{1}{c}{\checkmark} \\
	RE &
	  \multicolumn{1}{c}{\checkmark} &
	  \multicolumn{1}{c}{\checkmark} &
	  \multicolumn{1}{c}{\checkmark} &
	  \multicolumn{1}{c}{\checkmark} \\
	{EA} &
	  \multicolumn{1}{c}{x} &
	  \multicolumn{1}{c}{x} &
	  \multicolumn{1}{c}{x} &
	  \multicolumn{1}{c}{x} \\
	{NLP} &
	  \multicolumn{1}{c}{x} &
	  \multicolumn{1}{c}{x} &
	  \multicolumn{1}{c}{x} &
	  \multicolumn{1}{c}{x} \\
	  \arrayrulecolor{black}
	  \bottomrule
	\end{tabular}
	}
	\\[0.2cm]
	\textbf{RE}: regular expressions,
	\textbf{FOL-Exp.}: first-order logic expressiveness, \\
	\textbf{EA}: external access,
	\textbf{NLP}: natural language processing
\end{table*}

%% file: text/4_concept.tex
\section{Approach}
\label{sec_approach}

In our work with experts in the cultural heritage domain, we found that these experts typically have limited knowledge of database technologies.
Consequently, they need the IT support from data engineers to perform domain-specific data quality analysis, which can result in time-consuming and error-prone workflows.

Our analysis of popular database technologies revealed significant similarities in their query languages (\autoref{sec_analysis}).
Furthermore, many data quality problems have similar structures (see \autoref{sec_motivation}).
We concluded that a template-based approach to domain-specific data quality analysis is promising.

With this in mind, we present the Quality Pattern Model framework (QPM).
This framework allows templates to be defined generically, i.e., independently of any specific database technology.
With QPM, domain experts should be able to define domain-specific data quality analysis without extensive technical knowledge.
The overarching goal of QPM is to provide an accessible framework that enables domain experts to define quality analyses while reducing the necessary expertise in query languages, database technologies, and formats.
Furthermore, QPM supports the specification of domain-specific data analyses in a technology-independent manner, so they need only be created once and can be instantiated across different database technologies as needed.

\input{text/4b_workflow}

\input{text/4a_concept}

\subsection{Example}\label{subsec_example}

In this section, we illustrate the approach presented above using the example presented in \autoref{sec_example}.
To check for completeness violations, we search for \textit{all painting nodes in the database that do not have a link to an artist node}.
In \autoref{subsec_querycomparison}, we presented two queries in XQuery and Cypher for this specific data quality problem.
The logical structure behind these queries is similar.
Both search for paintings, for which the specification of an associated artist is missing.
In abstract terms, they \emph{search for} nodes of a specific type \emph{under the condition} that they have no relationships with nodes of another specific type.

\begin{figure*}
	\centering
	\begin{subfigure}[t]{.325\linewidth}
		\centering
		\includegraphics[width=\linewidth]{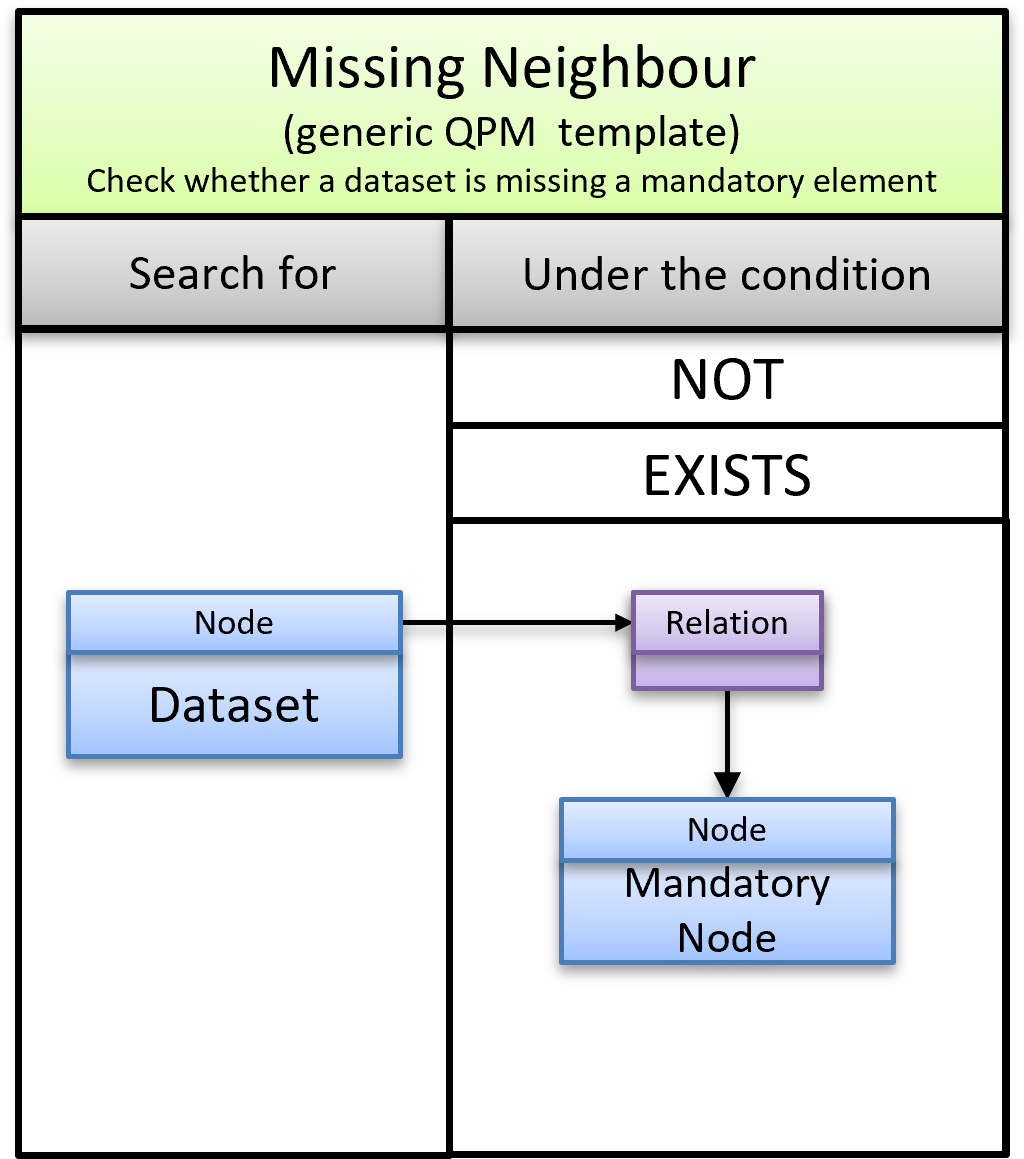}
		\caption{Generic QPM template}
		\label{fig_missing_generic}
	\end{subfigure}
	\begin{subfigure}[t]{.325\linewidth}
		\centering
		\includegraphics[width=\linewidth]{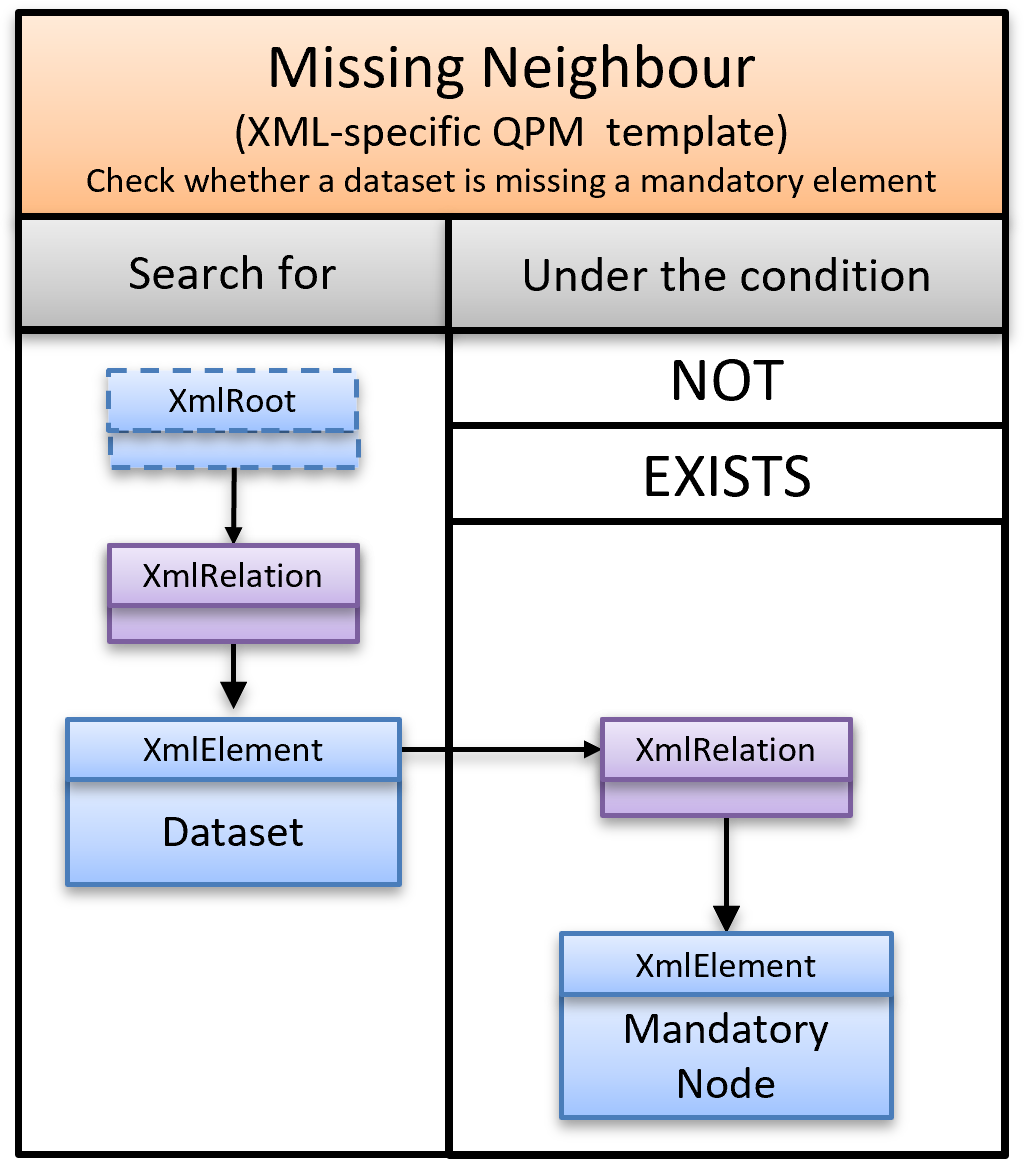}
		\caption{XML-specific QPM template}
		\label{fig_missing_xmlabstract}
	\end{subfigure}
	\begin{subfigure}[t]{.325\linewidth}
		\centering
		\includegraphics[width=\linewidth]{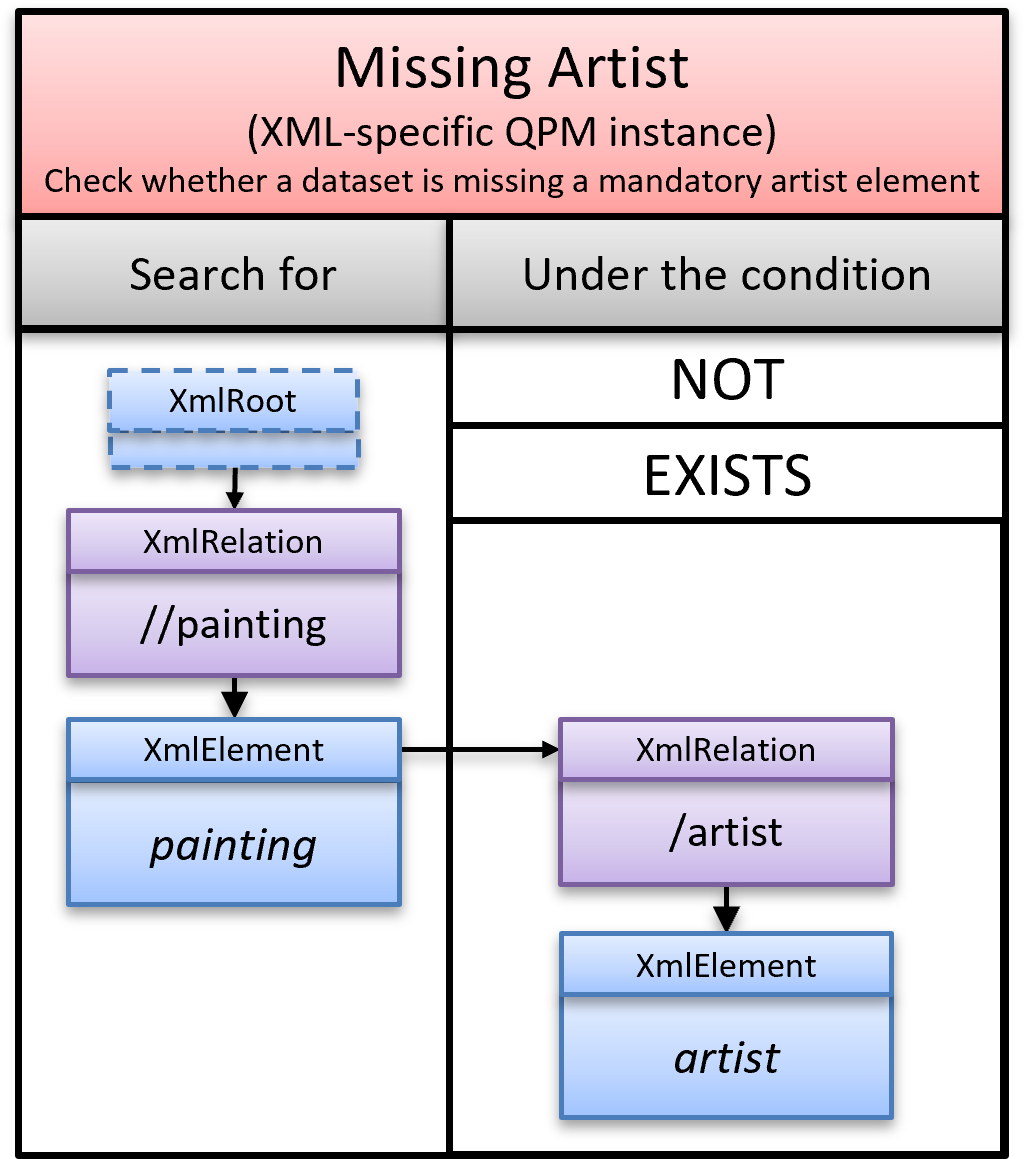}
		\caption{XML-specific QPM instance for the query in \autoref{lst_example-xml}}
		\label{fig_missing_xmlconcrete}
	\end{subfigure}
	\vspace{-0.3cm}
	\caption{QPM templates and instance for the running example in \autoref{sec_example}}
	\label{fig_missing_example}
\end{figure*}

Next, we show how these queries are specified using our workflow for the technology XML.
The generic QPM template in \autoref{fig_missing_generic} specifies a suitable abstract query structure.
It consists of a \emph{search for} part, which specifies the query result, and an \emph{under the condition} part, which formulates the condition to be checked. 

For this problem, a domain expert selects this generic QPM template and instantiates it for XML.
The result is an XML-specific QPM template (see \autoref{fig_missing_xmlabstract}).
During the automatic adaptation process, an XML root element is added, and all nodes and relations are made XML-specific.
 
In the next step, the domain expert adapts this XML-specific QPM template to a particular database schema.
The concrete adaptation with respect to our running example is shown in \autoref{fig_missing_xmlconcrete}.
This QPM instance specifies \emph{painting} as the XmlElement that plays the role of a dataset, and \emph{artist} as the XmlElement that plays the role of the mandatory node, as well as a suitable XmlRelation.
As a result, this QPM instance searches for all nodes of type {\sf painting} without an associated {\sf artist} in XML data.
Once the QPM instance is specified, it can then be automatically translated into the XQuery expression shown in \autoref{lst_example-xml}.

%% file: text/4b_workflow.tex
\subsection{Workflow}
\label{subsec_workflow}

\begin{figure}
	\centering
	\includegraphics[width=0.8\linewidth]{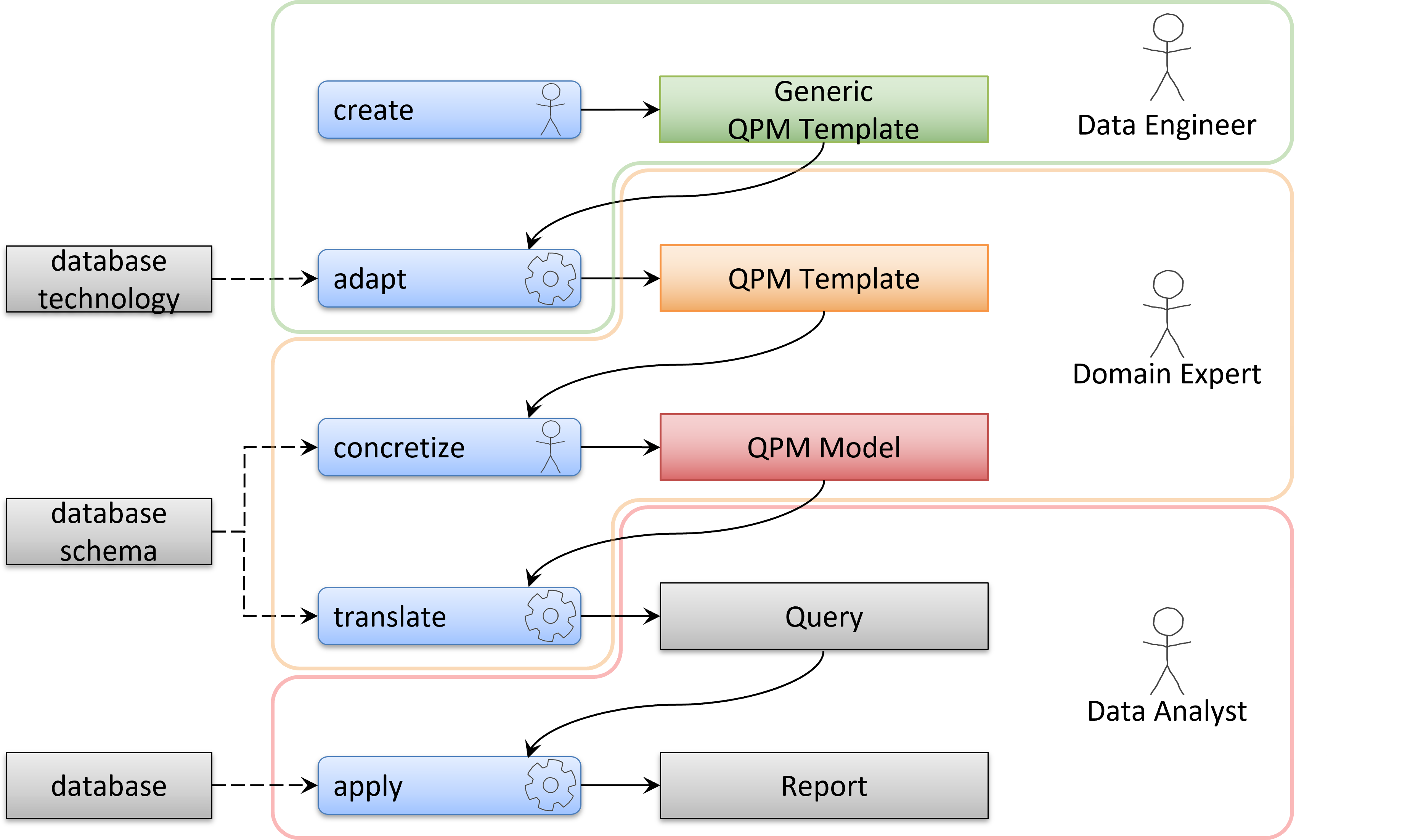}
	\caption{Workflow for the creation and application of QPM templates}
	\label{fig_workflow}
\end{figure}

We propose the following workflow for the template-based generation and application of domain-specific queries.
The workflow consists of three main use cases that are executed by different roles.
These three use cases are \emph{'Template Creation'}, \emph{'Analysis Definition'} (based on previously defined templates) and \emph{'Quality Analysis'} (using defined analyses).
Note that these use cases are not usually executed sequentially; however, the order above indicates their causal order.
E.g., the Template Creation use case only needs to be executed, if no fitting template exists yet for operationalization of a novel requirement.
The workflow is visualized in \autoref{fig_workflow}.

\paragraph{Use Case: Template Creation}

In this use case, a {\em data engineer} creates a {\em generic QPM template}.
A generic QPM template specifies a reusable query logic for quality analyses tailored to a specific type of data quality constraints independent of the database technology.
Quality analyses specified by an instantiated template can detect specific constraint violations and, consequently, quality problems.

Comprehensive domain-specific quality analyses require identifying various types of quality problems, so a variety of generic QPM templates are necessary.
New generic QPM templates need to be created by data engineers only if a novel kind of quality requirement occurs, minimizing the need for additional generic QPM templates over time.

Generic QPM templates are automatically adapted to specific database technologies, such as XML, becoming technology-specific QPM templates.
For this purpose, technology-specific parameters are integrated into the structure of a generic QPM template.
These parameters specify the data nodes and relationships within the selected technology. 
The result can be stored in a QPM template library.
Starting with a prebuilt library of QPM templates (e.g. \cite{zenodo-templatesset}) can reduce initial work.

To create generic QPM templates, data engineers must have a comprehensive understanding of first-order logic and data organization.
In addition, they must be able to think abstractly, structurally, and analytically.

\paragraph{Use Case: Analysis Definition}

This use case is intended for {\em domain experts} (e.g. cultural historians) who want to identify domain-specific data quality problems without requiring technical knowledge.
To specify a quality analysis, domain experts select a QPM template from the template library.
The selection process requires only an informal understanding of the QPM templates.
If no fitting QPM template exists, a data engineer needs to be consulted to initiate the Template Creation use case.
As long as the parameters are not set, a QPM template remains independent of a specific database schema (such as an XML schema).
Consequently, QPM templates cannot yet be applied to a database.

Next, the domain experts specify how the QPM templates are concretized to a selected domain-specific database schema.
Each parameter of the QPM template must be set, fitting the database schema to become executable.
During this process, users must specify the values of all parameters as well as the metadata for the QPM template, such as its name and description.
The result is a {\em QPM instance}, which represents a specific data quality constraint.
A QPM instance is ready to analyze databases of a specific format by compiling database-specific queries to detect these types of quality problems.
Rather than creating a new quality analysis, an existing QPM instance can be edited or duplicated.

Concretizing a QPM template requires an understanding of the database schema and relevant domain knowledge.
Based on the selected database schema and parameters, assisting information can be provided to minimize the required technical knowledge, such as suggesting values for specific data fields.
To create a comprehensive, domain-specific quality analysis for a specific database schema, a set of QPM instances can be created and stored in a library of QPM instances.

\paragraph{Use Case: Quality Analysis}

In this use case, a {\em data analyst} applies a domain-specific data quality analysis to a specific database.
The data analyst specifies the database to be analyzed and selects one or more constraints, which are represented by QPM instances from the QPM instance library.
The analysis searches for constraint violations, i.e. quality problems, in the data.

For the analysis, QPM automatically compiles a query in a traditional query language and applies it to the selected database.
The analysis result is a report of all matches of the compiled queries, i.e., a list of all constraint violations.
Based on this report, a data analyst can start the data improvement process.

%% file: text/4a_concept.tex
\subsection{Concept}
\label{subsec_concept}

\begin{figure*}[htbp]
	\centering
	\includegraphics[width=\linewidth]{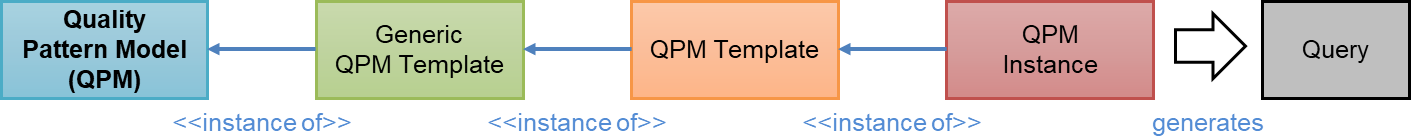}
	\caption{Concept structure}
	\label{fig_structure_approach}
\end{figure*}

To realize this proposed workflow, we have created a three-level modeling approach for domain-specific queries.
The three levels model the dependency between database technologies, database formats and concrete quality constraints.
This concept is visualized in \autoref{fig_structure_approach}.

{\em QPM} provides a modeling language for defining {\em generic QPM templates}.
This language can be used to create reusable query logic that can be customized and tailored to specific constraints.
Specifically, the generic QPM templates represent the logic of the quality analysis to identify certain types of constraint violations (cf. \autoref{sec_motivation}).
They abstract from the database technology and the database format (i.e. schema) and are therefore completely independent of technology-specific features.
A generic QPM template is instantiated to a {\em QPM template} by adapting it to a specific database technology.
infers parameters regarding the addressing of structure elements in relation to the root of an XML document.
To adapt a QPM template to a specific database schema, the parameters are set to concrete values.
The result is a fully parameterized {\em QPM instance} that is used to compile an analysis query in a selected query language.

%% file: text/5_design.tex
\section{Tool Support}
\label{sec_tool-support}

To validate the approach of QPM, we present a proof-of-concept tool implementing the workflow in \autoref{fig_structure_approach}.
The multi-layer modeling approach forms the backend of a web application and is implemented based on the Eclipse Modeling Framework\footnote{\url{https://projects.eclipse.org/projects/modeling.emf.emf} (2026-06-25)} (EMF), a quasi-standard framework for model-driven development.
The tool currently supports the database technologies XML, RDF, and Neo4j.
The backend provides a RESTful web API to communicate with other components, especially the user interface.
For the two main use cases of the tool, namely the definition and the application of domain-specific data quality analysis, there is a web frontend called \lstinline|Constrainify|.
The entire tool, including the Constrainify frontend, can be found on Zenodo \cite{zenodo-constrainify} and GitLab\footnote{\url{https://gitlab.gwdg.de/aqinda/constrainify} (2026-06-25)}.
The implementation of the QPM backend is available on Zenodo \cite{zenodo-qpm} and GitHub\footnote{\url{https://github.com/Project-KONDA/pattern-based-quality-analysis} (2026-06-25)}.

\input{text/5a_architecture}
\input{text/5c_metamodel}
\input{text/5b_api}
\input{text/5d_custom_operator}
\input{text/5e_constrainify}

%% file: text/5a_architecture.tex
\subsection{Architecture}
\label{subsec_architecture}

\begin{figure}
	\centering
	\includegraphics[width=\linewidth]{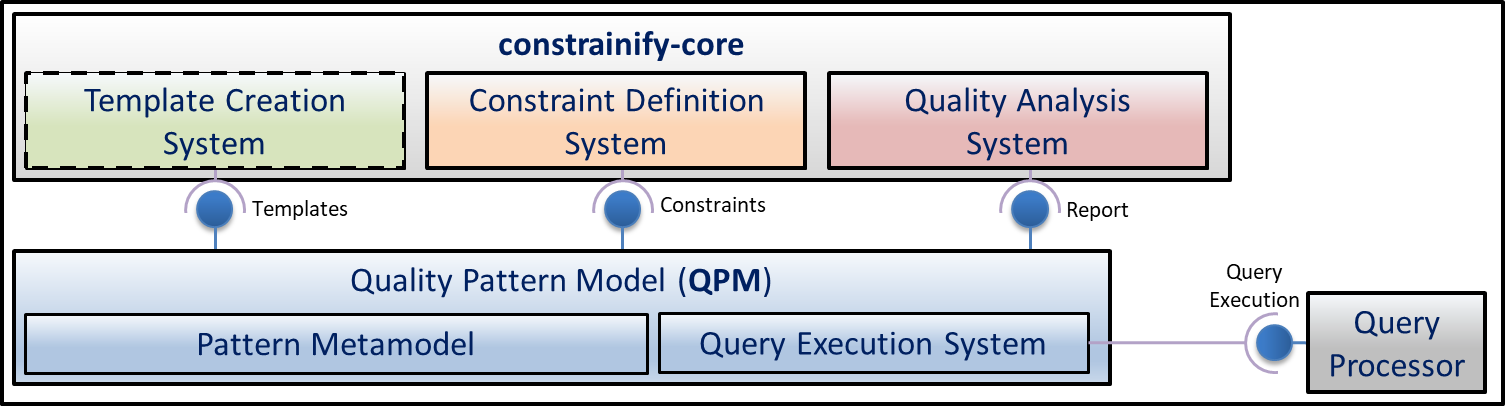}
	\caption{System architecture of the QPM Framework}
	\label{fig_components}
\end{figure}

\autoref{fig_components} provides an overview of the system architecture of our application Constrainify.
The core of the architecture is the \lstinline|Quality Pattern Model| (\emph{QPM}).
This component manages and stores predefined and user-defined QPM templates and QPM instances in a library.
Its \lstinline|Pattern Metamodel| defines the abstract structure and supported functionalities of the (generic) QPM templates and QPM instances.
QPM also includes the \lstinline|Query Execution System|.
It is used to identify constraint violations, by executing the generated queries of QPM via established query processors (e.g. Saxon\footnote{\url{https://www.saxonica.com} (2026-12-18)} for XML) against the provided database.

The components at the top of \autoref{fig_components} form the frontend components of our architecture, called \lstinline|constrainify-core|.
They interact with the QPM backend exclusively through APIs to ensure loose coupling, consistency, and maintainability.
The frontend is structured in three components, corresponding to the use cases in \autoref{subsec_workflow}:

The \lstinline|Template Creation System| is designed as a graphical modeling workbench for defining new generic QPM templates.
A concrete example is shown in \autoref{fig_missing_generic}.
All generic QPM templates can be automatically adapted to a specified database technology to become technology-specific QPM templates.
See \autoref{fig_missing_xmlabstract} for a concrete example.
This graphical modeling workbench currently only exists as a rough prototype.
The \lstinline|Analysis Definition System| provides a form-based view for entering parameter values into the adapted QPM template.
This results in a fully specified QPM instance.
See \autoref{fig_missing_xmlconcrete} for a concrete example.
The \lstinline|Quality Analysis System| enables users to upload data and select a subset of valid QPM instances.
This component uses the \lstinline|Query Execution System| for the specified quality analysis on the uploaded data.
Identified quality problems are presented to the user in an analysis report.

%% file: text/5c_metamodel.tex
\subsection{Quality Pattern Metamodel}
\label{subsec_metamodel}

At the core of QPM is the metamodel that defines the abstract structure of quality analyses using a Java class structure.
This metamodel implements how different (generic) QPM templates and QPM instances can be built (such as the examples in the figures \ref{fig_missing_generic}, \ref{fig_missing_xmlabstract}, and \ref{fig_missing_xmlconcrete}).
It also defines their functionalities, including adaptation to technologies, parameterization, and translation into technology-specific query languages, namely XQuery, SPARQL and Cypher.
Note that the adaptation to relational databases is planned but not yet implemented.

\begin{figure}
	\centering
	\includegraphics[width=0.8\linewidth]{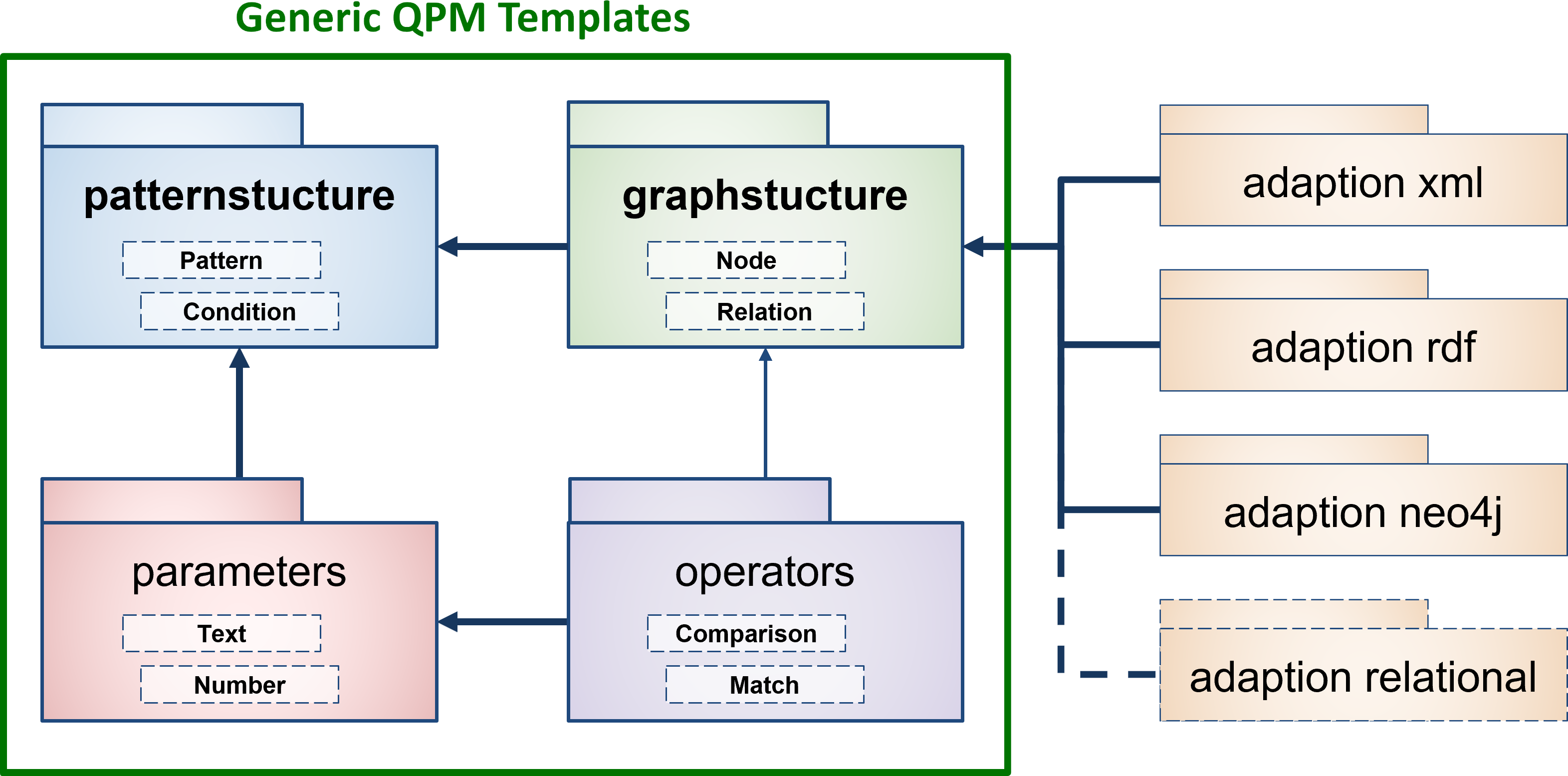}
	\caption{Package structure of the QPM metamodel including dependencies and example classes}
	\label{fig_overview}
\end{figure}

\paragraph{QPM Metamodel}
The metamodel of QPM is split across four main packages, as illustrated in \autoref{fig_overview}: 
The \lstinline|patternstructure| package deals with the definition of the formal logic of QPM templates.
It contains the general structure, quantifiers and formulas of the quality analyses.
The \lstinline|graphstructure| package is used to specify graphs that map to specific data substructures, such as nodes and relations.
The operators specified in the \lstinline|operators| package are used to define additional conditions within QPM templates, such as regular expression matching predicates.
Finally, the \lstinline|parameters| package defines the types of parameters that enable the concretization of QPM templates to QPM instances.

The packages for technology-specific adaptations are shown on the right side of \autoref{fig_overview}.
These packages provide the functionality to adapt the generic QPM templates to the particularities of each database technology, primarily the different database structures and navigational operators.

\paragraph{QPM Language}

The QPM metamodel defines the abstract syntax for database queries across multiple levels of abstraction.
According to Kleppe \cite{dsl}, a domain-specific language requires an abstract syntax, a concrete syntax, and semantics.
The QPM metamodel defines the abstract syntax.
The concrete syntax is a graphical one as shown in e.g. \autoref{fig_missing_example}.
The semantics are given via interpretability as executable quality analyses.

The QPM language is organized into multiple layers.
The technology-independent core, as defined by the four packages marked as generic in \autoref{fig_overview}, defines the generic QPM language.
This core comprises generic, reusable data analysis templates.
The technology-specific packages extend this core to form technology-specific abstract, parameterized QPM languages.
By binding the parameters, fully specified QPM instances are obtained.
The set of all well-formed QPM instances over a given technology forms a concrete, technology-specific QPM language.

%% file: text/5b_api.tex
\subsection{RESTful Web API}
\label{subsec_api}

To facilitate communication with other components, especially the user interface (cf. \autoref{fig_components}), we implemented a REST API in the OpenAPI format\footnote{\url{https://raw.githubusercontent.com/Project-KONDA/pattern-based-quality-analysis/refs/heads/main/qualitypatternmodel/openapi.yaml} (2026-06-25)}.
The API allows specifying QPM templates with parameters and executing the resulting queries.
For this purpose, the API can manage multiple libraries of QPM templates, following the specified workflow (cf. \autoref{fig_workflow}).
One library contains all \emph{generic QPM templates}.
For each supported database technology, it handles a library of technology-specific \emph{QPM templates} and specified \emph{QPM instances}.
The libraries can be accessed with different retrieval criteria (filters and groups) to efficiently narrow down results.

For the use case \emph{'Analysis Definition'}, the API supports the creation of new analyses by instantiating QPM templates.
It also enables the loading, duplicating, modifying and deleting of existing analyses.
Editing and resetting parameter values and metadata of analyses is also possible.
Metadata includes the name, description, tags and an optional mapping to specific data formats or databases.
Custom metadata can also be added.

For the use case \emph{'Quality Analysis'}, the API supports the selection of one or more domain-specific quality analyses from a library.
These analyses can either be compiled into queries or applied directly to selected data.
The analysis results, which represent identified quality problems, are returned in a structured report.

%% file: text/5d_custom_operator.tex
\subsection{Custom Operator Extension}
\label{sec_implementation_java}

Relying only on the expressiveness of query languages can cover a substantial amount of constraints through quality analyses.
However, query languages are limited to intrinsic operators for data analysis.
This does not include features such as data verification or spell checks.

\begin{figure*}[thbp]
	\small
	\centering
	\begin{minipage}[t]{.6\linewidth}
		\centering
		\includegraphics[width=\linewidth]{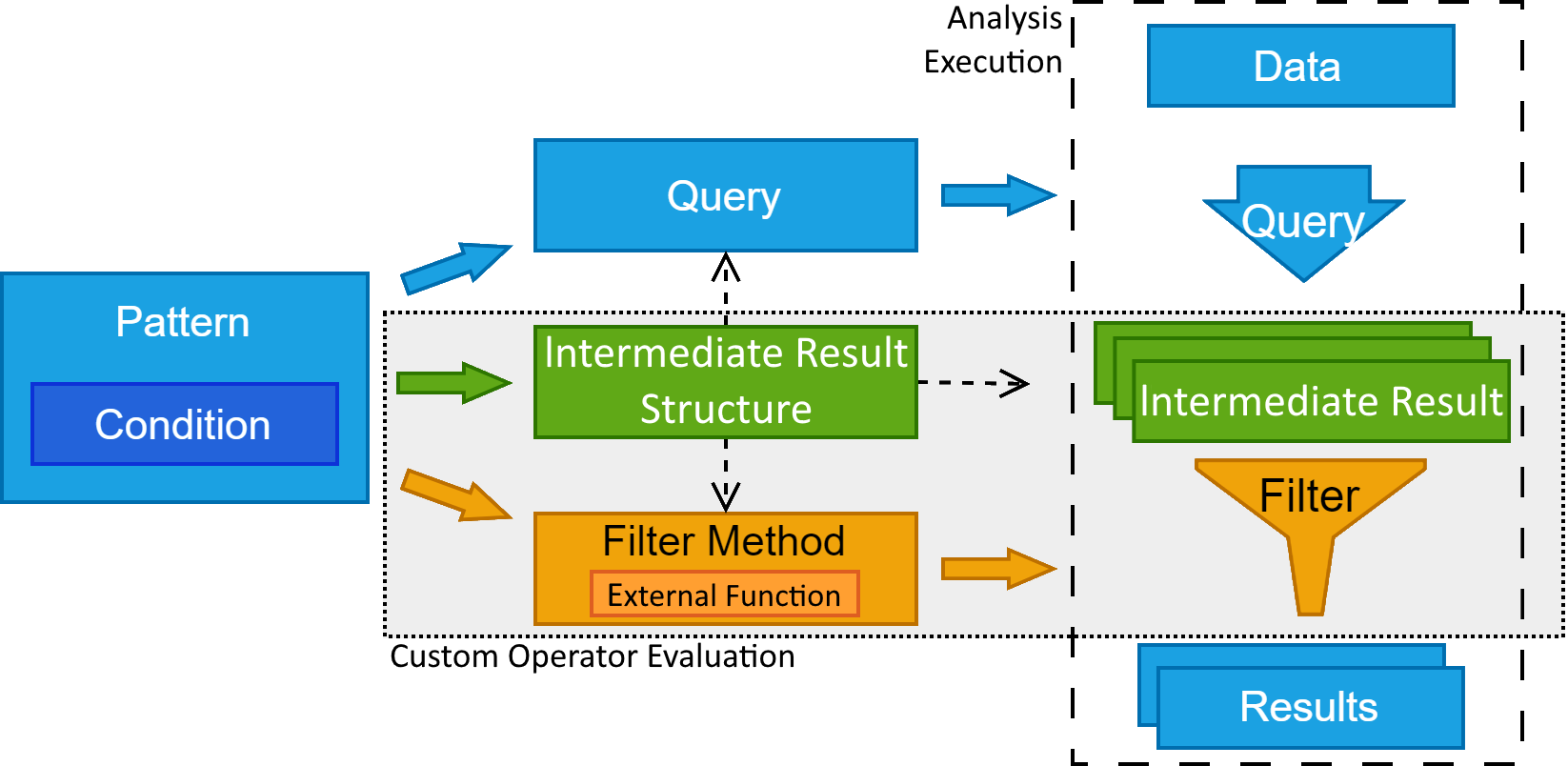}
		\caption{Concept for supporting custom operators in QPM by using externally programmed functions.}
	\label{fig_java-query_concept}
	\end{minipage}
	\begin{minipage}[t]{.04\linewidth}
	\end{minipage}
	\begin{minipage}[t]{.36\linewidth}
		\centering
		\includegraphics[width=.9\linewidth]{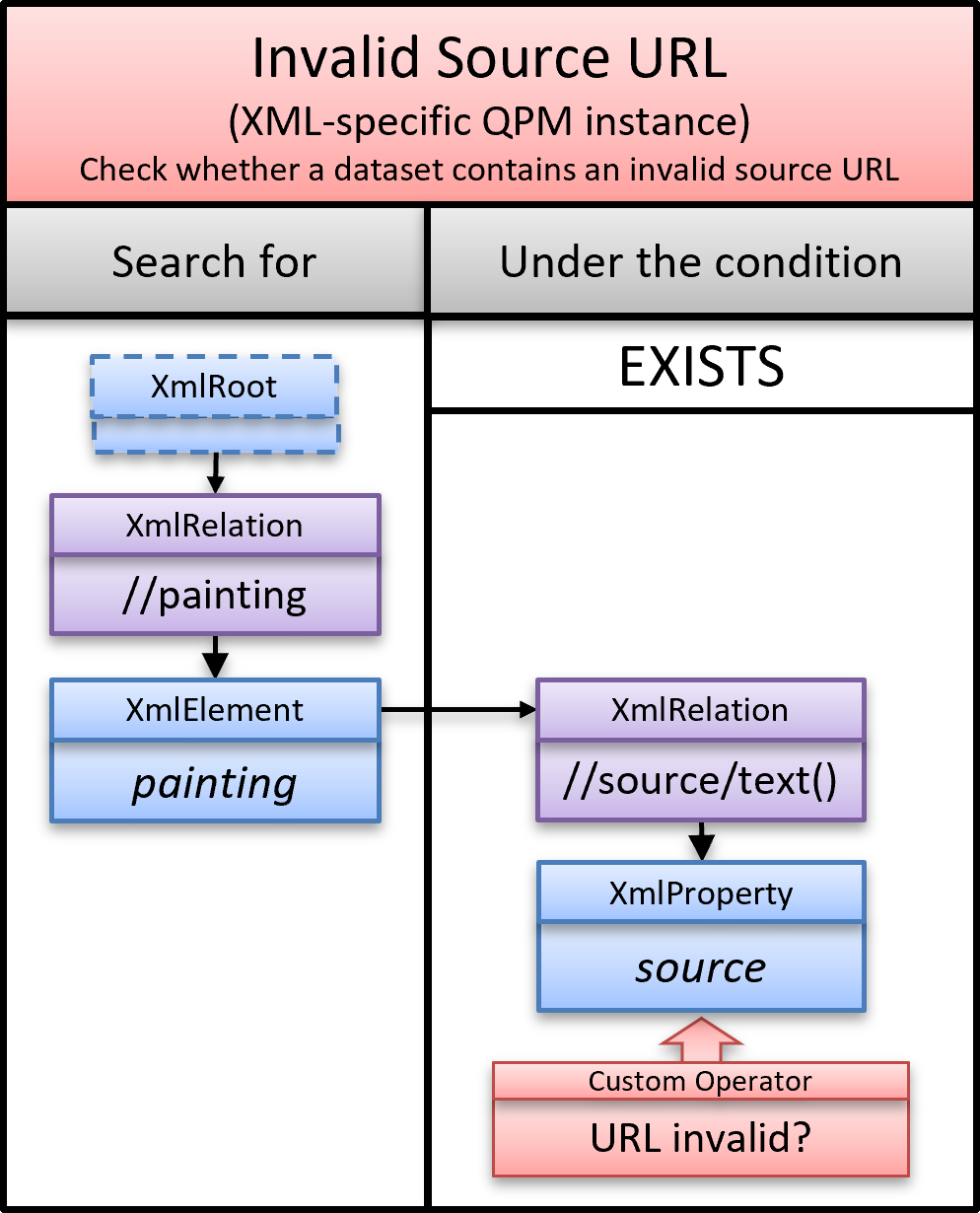}
		\caption{XML-specific QPM instance checking URL validity}
		\label{fig_valid-link}
	\end{minipage}
\end{figure*}

To enhance the expressiveness of QPM, we introduce custom operators into QPM templates based on functions written in a high-level programming language.
Thus, we extend the operator package of \autoref{fig_overview} with operators using custom functions.
In accordance with QPM, we have chosen Java as programming language for this tool extension.
As presented in \autoref{fig_java-query_concept}, the extension is based on evaluating as much of each QPM instance as possible by leveraging technology-specific query languages for predicate filtering, thereby maximizing database pushdown optimization~\cite{pushdown}.
The entire process is automated.

A quality analysis is generated from the QPM instance and evaluated to the extent that the expressiveness of the query languages allows, producing intermediate results.
These intermediate results contain all possible results, accompanied by values for indicating whether each result is true.
Finally, a filter is applied to the intermediate results.
This filter identifies the true results from the intermediate results by assessing the additional values.
This process uses external functions from higher-level languages.
Because the structure of the intermediate results must fit the expected filter input, they must also adhere to a defined, common structure.
Thus, a specific structure is generated for the quality analysis of the QPM instance.

We have proven this concept by implementing it using XQuery and Java.
This allows the database query to produce intermediate results that provide a $string$ as input to a Java method. 
The method must return a $boolean$ as output.
This enables quality analyses, such as searching for invalid URLs (as $string$) in a dataset, which goes beyond the capabilities of query languages.
This concept is also useful for verifying addresses, email addresses, and ISBN numbers, as well as language and spelling checks.
However, note that these features may require a network connection and external services, such as geocoding, registry lookup, and natural language processing.

Next, we will present an example of how to analyze XML data by validating all source URLs of the metadata present in a painting data record.
The goal is to check the validity of all URLs that substantiate the metadata of paintings in an XML dataset.
Given a suitable pattern, such as the one in \autoref{fig_valid-link}, QPM generates an intermediate result structure, an XQuery, and a filter method.
When analyzing a dataset, the generated query is executed first.
The query extracts all painting records from the dataset and complements each one with a list of all source URLs of the record.
The query output structure combines these into a small XML structure that conforms to the specified intermediate result structure.
The filter then iterates over all intermediate results and evaluates all URLs.
If an invalid URL is identified, the associated painting record is flagged as a data quality problem.
The identified painting records are then returned as quality problems in the quality report.

%% file: text/5e_constrainify.tex
\subsection{Web Frontend \lstinline|Constrainify| } 
\label{sec_implementation_gui}

While data engineers possess the skills to define templates using code, domain experts and data analysts need a user interface (UI) to be able to define and execute data quality analyses.
For this we provide a web UI \lstinline|Constrainify|~\cite{constrainify_whitepaper}.
\lstinline|Constrainify| implements the frontend components \lstinline|Analysis Definition System| and \lstinline|Quality Analysis System| presented in \autoref{subsec_architecture}.

Given a set of QPM templates and QPM instances, \lstinline|Constrainify| currently supports two of the three use cases of QPM described in \autoref{subsec_workflow}.
The \lstinline|Quality Analysis System| implements the primary use case \emph{'Quality Analysis'}, where a data analyst can select one or more QPM instances, which represent predefined quality analyses to analyze a given database.
This use case is complemented by the use case \emph{'Analysis Definition'}, implemented in the \lstinline|Quality Analysis System|, where new QPM instances can be specified from QPM templates.
Currently, as long as the \lstinline|Template Creation System| is lacking, QPM also provides an extensive list of pre-defined XML-specific QPM templates \cite{zenodo-templatesset}.
Currently, Constrainify is limited to analyses for the technology XML, since XML is a common harvesting format for institutions, such as those in the cultural heritage domain.
Extending Constrainify to other database technologies is left to future work.
We expect this to be straightforwardly achievable.
The missing support relates to modules that specify technology-specific parameters, namely paths and node types.

The frontend Constrainify presents the quality analyses as constraints (see \autoref{sec_example}), defined by QPM templates and QPM instances.
This is important because domain experts think in terms of quality requirements instead of quality problems.
It allows them to focus on the expected state of the data rather than how to detect data quality problems.

\paragraph{Use Case: Quality Analysis}
\begin{figure}
	\centering
	\includegraphics[width=0.7\linewidth]{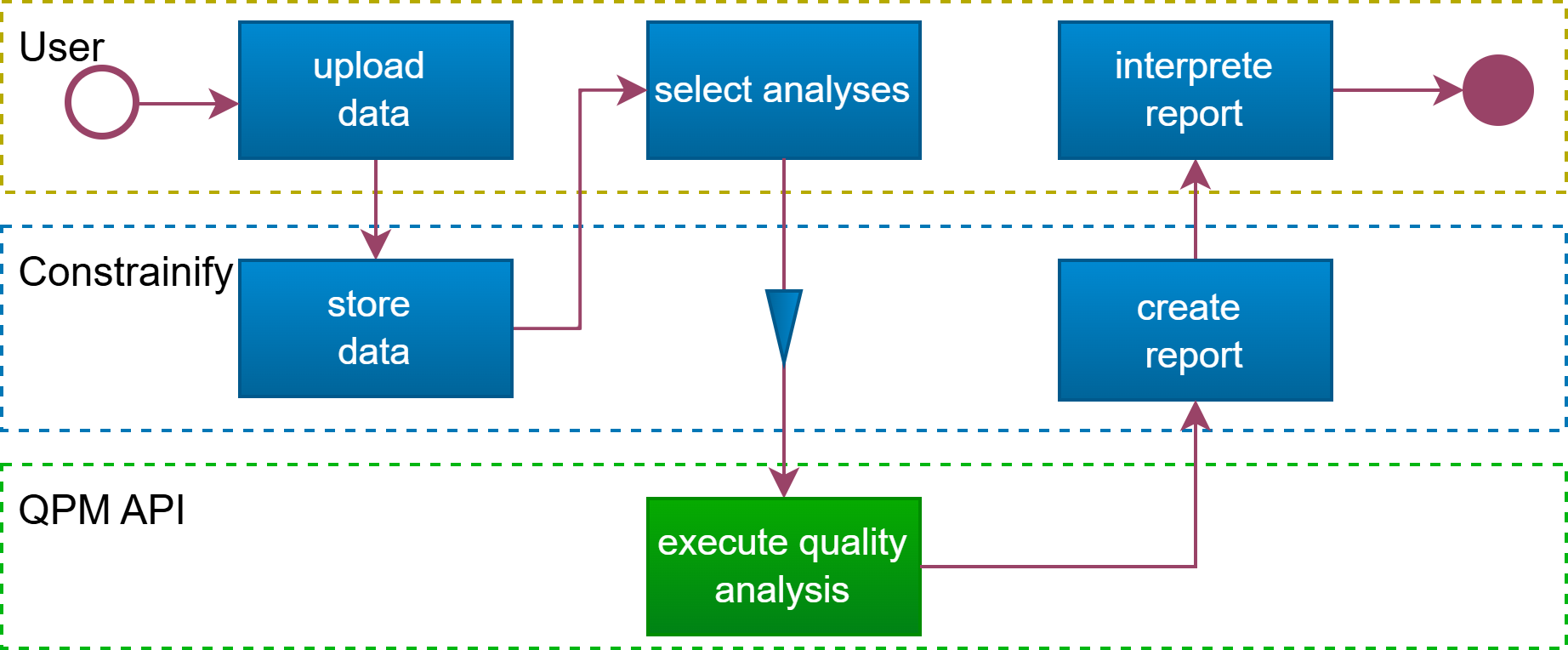}
	\caption{Use Case: Quality analysis by a data analyst}
	\label{fig_constrainify_uc1}
\end{figure}

\begin{figure}
	\centering
	\includegraphics[width=0.8 \linewidth]{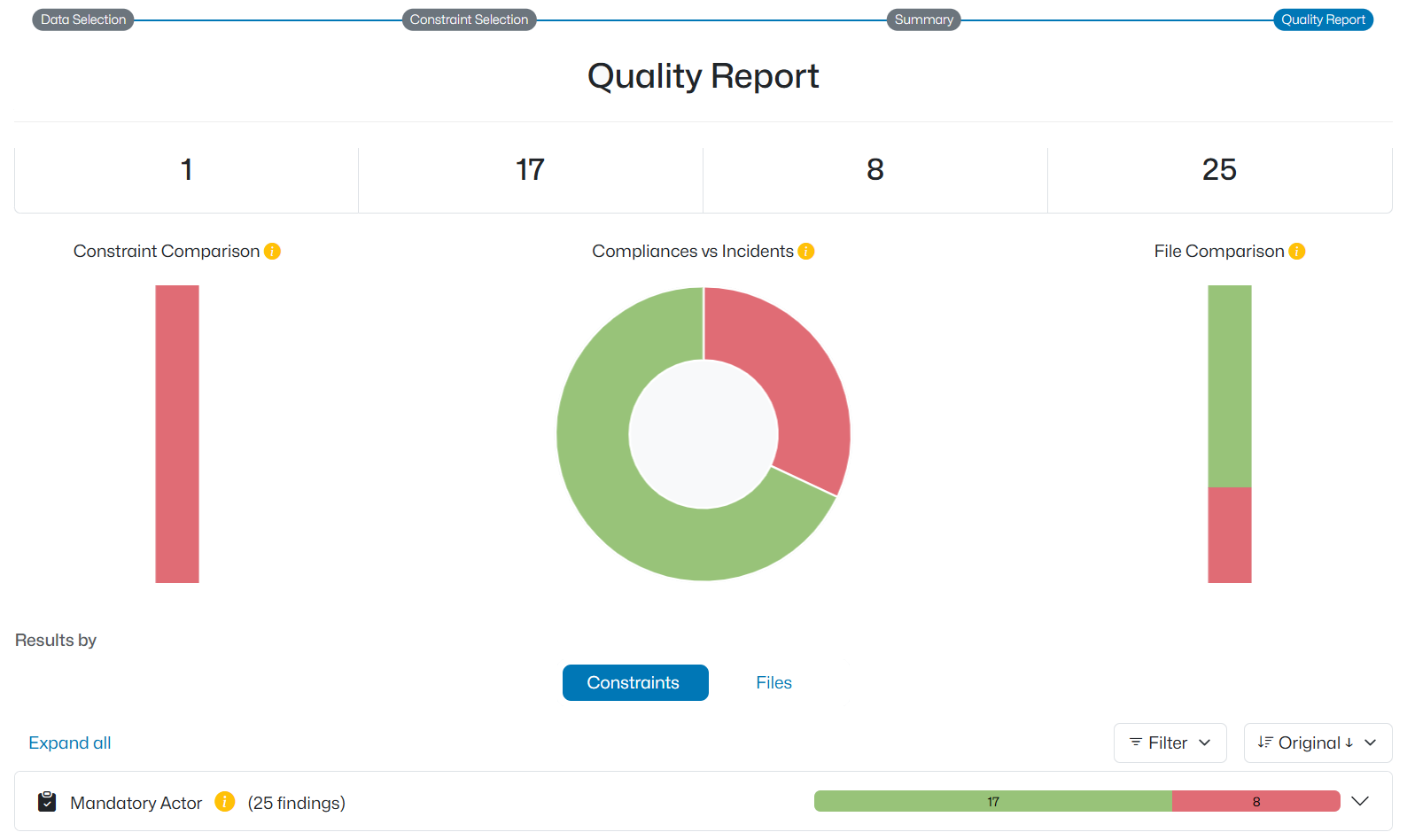}
	\caption{Screenshot of a quality report in \lstinline|Constrainify|}
	\label{fig_constrainify_report}
\end{figure}

\autoref{fig_constrainify_uc1} illustrates the workflow for the primary use case of performing a quality analysis.
A data analyst begins by uploading their data to the \lstinline|Constrainify| frontend.
Based on the uploaded data, the data analyst is presented with a list of available constraints that match the database technology and data model.
Each constraint represents a QPM instance in the QPM backend that specifies a quality analysis.
The data analyst must select at least one constraint.
Once the selection has been made, \lstinline|Constrainify| initializes a data analysis by the QPM backend.
\lstinline|Constrainify| can then present the results in the form of a quality report.
This quality report shows which constraints are violated and where the violations occur in the data.
The data analyst can then interpret the results and start improving their data.
\autoref{fig_constrainify_report} presents an example report.

\paragraph{Use Case: Analysis Definition}
\begin{figure}
	\centering
	\includegraphics[width=\linewidth]{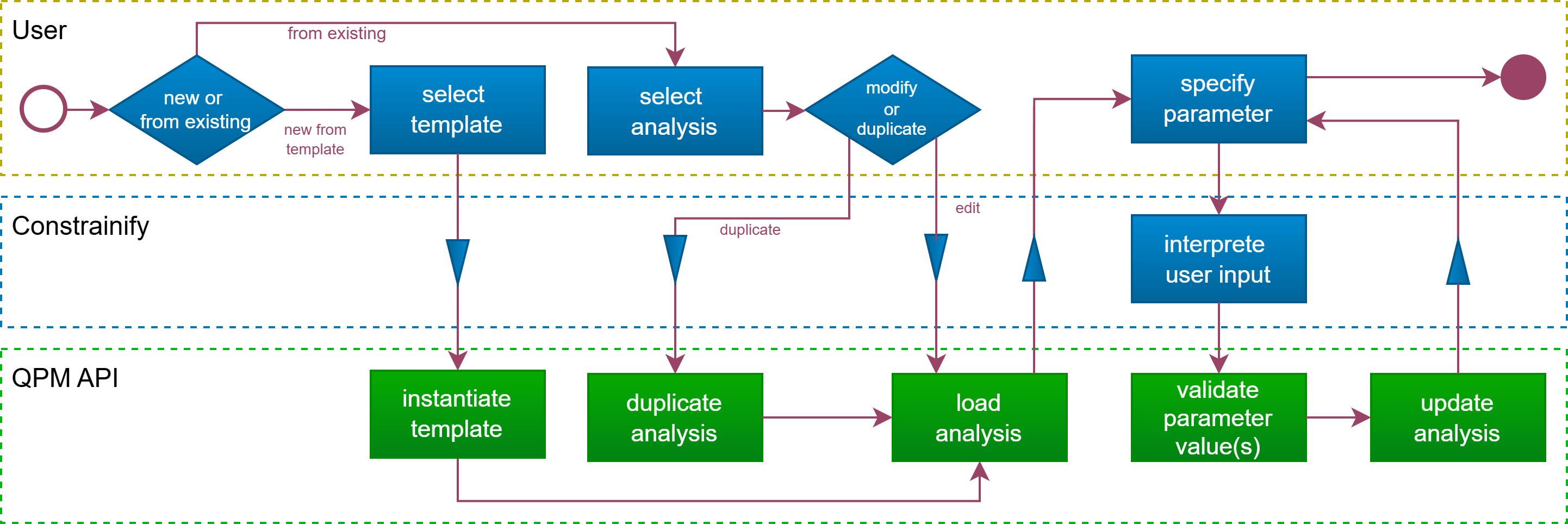}
	\caption{Use Case: Analysis definition by a domain expert}
	\label{fig_constrainify_uc2}
\end{figure}

\begin{figure}
	\centering
	\includegraphics[width=0.8 \linewidth]{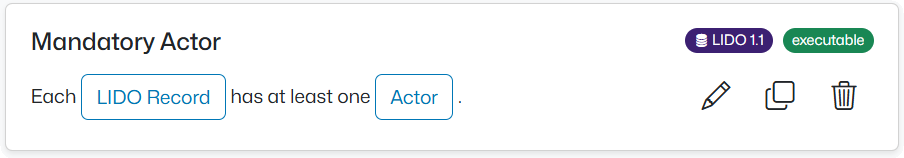}
	\caption{Screenshot of a constraint in \lstinline|Constrainify|}
	\label{fig_constrainify_constraint}
\end{figure}

\begin{figure}
	\centering
	\includegraphics[width=0.8 \linewidth]{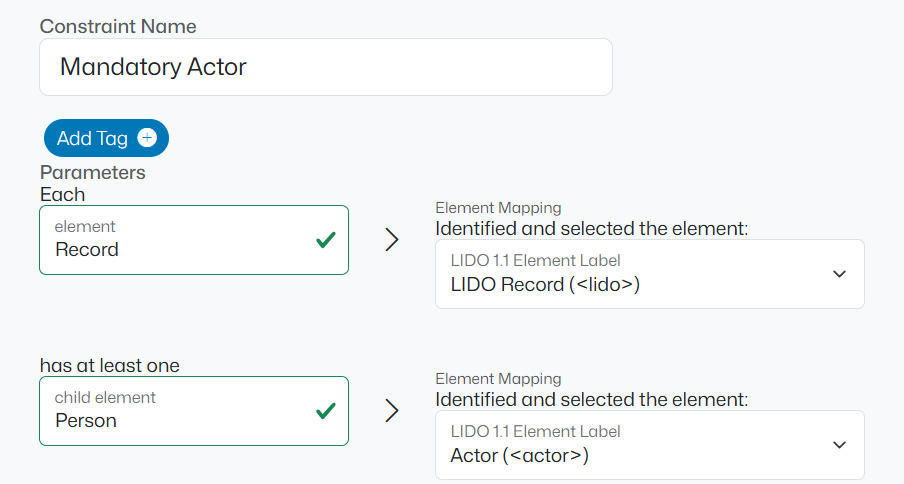}
	\caption{Screenshot of a parameterization view in \lstinline|Constrainify|}
	\label{fig_constrainify_parameterization}
\end{figure}

The secondary use case involves a domain expert specifying a constraint for a data analysis.
This workflow is represented by \autoref{fig_constrainify_uc2}.
\lstinline|Constrainify| provides three options for specifying a new constraint in the form of a QPM instance:
editing or duplicating an existing QPM instance or creating a new QPM instance based on a QPM template.
When creating a constraint, \lstinline|Constrainify| presents a list of available QPM instances to instantiate.

\lstinline|Constrainify| provides supporting features to abstract from technical details.
For example, users do not need to fully understand the underlying data model.
Instead, a natural language input is possible, which is automatically interpreted and mapped to technical parameters.
QPM then validates the specified parameters and updates the quality analysis accordingly.
An example constraint similar to the one in our running example for the Lido 1.1 data model is shown in \autoref{fig_constrainify_constraint}.
This example has two parameters to set, a \q{LIDO Record} and an \q{Actor}.
In the parameterization view (in \autoref{fig_constrainify_parameterization}), these parameters can be specified by a domain expert using natural language.
For example, the expert could use the words \q{person} or \q{painter} to search for the LIDO field with the name \q{actor}.
Constrainify then maps the entries to the appropriate fields of the selected data format, here LIDO.

%% file: text/6_evaluation.tex
\section{Evaluation}\label{sec_evaluation}

The general goal of our model-driven approach QPM is {\em to enable domain experts to define domain-specific data quality analyses in a technology-independent manner.
This eliminates the need for database technology expertise and avoids defining quality analyses several times for different database technologies.}
To achieve this goal, the approach must be comprehensive to ensure its applicability in real-world scenarios and its practical usability for domain experts.
Comprehensibility encompasses soundness, applicability, and expressiveness.
The \emph{soundness} of the implementation is guaranteed by a broad test suite of JUnit tests.
Since QPM has been implemented for the database technologies XML, RDF, and Neo4j, we conducted an evaluation that demonstrates the broad \emph{applicability} of QPM through real-world case studies using these technologies.
We compared the expressiveness of QPM with that of the traditional query languages, namely XQuery, SPARQL, and Cypher.
Furthermore, we evaluated expressiveness in terms of data quality problems collected from practice that can be addressed (cf. \autoref{sec_quality-problems}).
Finally, to demonstrate the \emph{usability}, we conducted a qualitative user study with the \lstinline|Constrainify| frontend.
This raises the following research questions:

\begin{enumerate}[leftmargin]
	\item[\textbf{RQ1:}] To what extent is QPM applicable in practice? (\autoref{sec_evaluation_rq1})
	\item[\textbf{RQ2:}] How expressive is the quality analysis language of QPM? (\autoref{sec_evaluation_rq3})

	\begin{enumerate}
		\item[\textbf{a)}] How expressive is the QPM language compared to other query languages?
		\item[\textbf{b)}] How expressive is the QPM language for data problem detection?
	\end{enumerate}
	\item[\textbf{RQ3:}] Given the Constrainify user interface, how usable is QPM for domain experts? (\autoref{sec_evaluation_rq4})
\end{enumerate}

\input{text/6_rq1_applicability}

\input{text/6_rq3_expressiveness}
\input{text/6_rq4_useability}

%% file: text/6_rq1_applicability.tex
\subsection{Applicability in Practice (RQ1)}
\label{sec_evaluation_rq1}

To demonstrate the applicability of QPM in practice, we tested two scenarios:
In \autoref{sec_evaluation_rq1a}, we select an existing data quality requirement document from practice and use it as basis to define templates for data quality analysis.
The result is a set of 30 generic QPM templates.
We instantiated the templates accordingly to create a set of QPM instances to cover the requirements catalog.
In \autoref{sec_evaluation_rq1b}, we tested the technology-independent applicability of 30 generic QPM templates, by applying them to three large collections in the three QPM-supported database technologies:
XML, RDF, and Neo4j.

\subsubsection{Use-case-specific Applicability}
\label{sec_evaluation_rq1a}

Our project partner, the Deutsche Digitale Bibliothek (DDB), collects data from various institutions, including museums, media libraries, and heritage preservation organizations, and makes it available online.
This delivered data mainly uses the LIDO XML format.
The DDB provided us with multiple documents that outline requirements for data quality in natural language and define quality guidelines for data deliveries.
These requirements were collected manually during the DDB's continuous process of integrating data into its online portal.
The requirements are specific to the LIDO\footnote{``Lightweight Information Describing Objects'' \url{https://icom-documentation.mini.icom.museum/working-groups/lido/lido-overview/about-lido/what-is-lido/} (2026-06-25)} XML format.
Specifically, the documents contained requirements for delivery data\footnote{\url{https://deutsche-digitale-bibliothek.atlassian.net/wiki/spaces/DFD/pages/48103977/Anforderungen+an+die+Lieferdaten} (2026-06-25)} and format-specific requirements for DDB-LIDO\footnote{\url{https://deutsche-digitale-bibliothek.atlassian.net/wiki/spaces/DFD/pages/48104132/DDB-LIDO} (2026-06-25)}.
Additionally, we received documents regarding the requirements for a minimum record\footnote{\url{https://www.minimaldatensatz.de/} (2026-06-25)} and from the MQ project\footnote{\url{http://ddb.qa-catalogue.eu/ddb-qa-2.0/?lang=en} (2026-06-25)}.

We formalized the requirements from these documents into a set of 130 data quality constraints.
This was done in close consultation with the DDB, resulting in some modifications and additional constraints.
Nine of the constraints were identified as 'out-of-scope' because they were formulated too abstractly to be checked computationally (e.g. 'the title must be meaningful') or because they described constraints targeting the system rather than the data (e.g. 'the identifier must not be changeable').
Based on the resulting natural language requirements, we developed 30 QPM templates that encompass different types of data quality problems and analysis techniques.
During development, we added additional QPM templates and variants to our template library and we will continue to do so.
An overview over the implemented pattern templates can be found in our repository\footnote{\label{pattern-overview}Pattern Overview \url{https://github.com/Project-KONDA/pattern-based-quality-analysis/blob/develop/qualitypatternmodel/src/qualitypatternmodel/newservlets/patterns/PatternOverview.xlsx} (2026-06-25)}.
While not all of these QPM templates were required to realize the constraint set, they seem useful for future applications. 

These 30 QPM templates were sufficient to realize 116 constraints as a set of QPM instances \cite{zenodo-constraintsset-d}, which specify 96\% of the requirements of the DDB documents.
Only five constraints remained unrealized due to the absence of two features:
the full integration of an external service and the ability to span analyses across multiple files.
We reviewed the defined QPM instances in detail with domain experts.
The resulting set of QPM instances enables a comprehensive analysis of multiple real-world datasets.

\subsubsection{Technology-independent Applicability}
\label{sec_evaluation_rq1b}

Secondly, we tested the applicability of the 30 generic QPM templates created to all supported database technologies.
We instantiated the generic QPM templates for three collections involving the three database technologies that QPM supports:
XML, RDF, and Neo4j.
To validate the technology-specific QPM templates for quality analyses, we instantiated them for the following real-world data collections.

The German Documentation Center for Art History – Bildarchiv Foto Marburg\footnote{\label{fotomarburg}\url{https://www.uni-marburg.de/de/fotomarburg} (2026-06-25)} provided us with an XML dataset in the LIDO format.
This cultural heritage data is a part of the largest image archive of European art and architecture.
The dataset contains over 700,000 data records of cultural heritage objects and over 47 million XML elements.
Since the DDK uses the same data format as the DDB, we could apply the constraints specified in \autoref{sec_evaluation_rq1a}.
For RDF, we used the public Wikidata\footnote{\url{https://www.wikidata.org} (2026-06-25)} database.
It is the largest open, collaborative knowledge base, governed by crowd-sourcing and currently contains over 120 million data items.
The Neo4j use case is covered by the Neo4j graph database, which is provided by the scholarly Regesta Imperii project\footnote{\url{http://www.regesta-imperii.de/en/} (2026-06-25)}, which contains approximately 145,000 regesta numbers describing chronologically ordered historical documents.
This is one of the first advancements to use a graph database for cultural heritage research.

\autoref{tab_application} shows the key metadata for each selected collection.
These data collections are representative of applications for QPM because they differ fundamentally in terms of technology, use, and governance.
Nevertheless, significant data quality problems are evident in all of them, underscoring the need for quality analysis tools.
While all 30 generic templates can be adapted to XML data, only 21 of them are applicable to RDF and Neo4j data.
The remaining templates rely on custom operators that are currently XML-exclusive (see \autoref{sec_implementation_java}).
We found at least one useful instance of each template for each dataset.

\input{text/6_table_applicability}

\subsubsection{Answer to RQ1}
\label{sec_evaluation_rq1c}

We demonstrated that QPM can be used to formulate templates for conducting comprehensive, domain-specific data quality analyses. 
We achieved 96\% of the constraints for XML data using 30 generic QPM templates.
70\% of the QPM templates were adaptable to all three database technologies: XML, RDF, and Neo4j.
The only limitation is the current state of implementation (see \autoref{sec_evaluation_rq1a}).
Representative real-world case studies show the practical applicability of QPM for domain-specific data quality analysis.

%% file: text/6_table_applicability.tex
\begin{table}[]
	\centering	
	\caption{Key data of real-world case studies using different database technologies with 30 generic templates}
	\label{tab_application}
	\begin{tabular}{@{}lllll@{}}
	\toprule
	            & DDK / DDB Lido & Wikidata & Regesta Imperii \\
	\midrule
	Technology & XML/XQuery & RDF/SPARQL        & Neo4j/Cypher \\
	Governance & institutional & scholarly-academic & crowd-sourced \\
	Database size       & 700,000 records & 120 million entities & 145,000 records \\
	\# Generic Templates   & 30        & 30       & 30              \\
	\# Abstract Templates   & 30        & 21       & 21              \\
	\# Constraints & 116        & 21       & 21              \\
	\bottomrule
	\end{tabular}
\end{table}

%% file: text/6_rq3_expressiveness.tex
\subsection{Expressiveness (RQ2)}
\label{sec_evaluation_rq3}

Our model-driven approach to defining domain-specific quality analyses is based on the assumption that common database query languages, such as XQuery\footref{xquery}, SPARQL\footref{sparql}, and Cypher\footref{cypher}, have similar levels of expressiveness and key features.
As shown in \autoref{subsec_dbstructure_comparison}, there are many similarities between common database technologies and their respective query languages.
The generic QPM templates defined with QPM can be translated into different query languages.
\autoref{sec_evaluation_quality_problems} investigates the expressiveness of QPM in addressing the quality problems identified in the cultural heritage domain (see \autoref{sec_quality-problems}).
In \autoref{sec_evaluation_query_languages}, we investigate the expressiveness of QPM in terms of the query language features that are supported, those that are feasible but will be addressed in the future, and those that are potentially feasible.

\input{text/6_rq3_expressiveness_languages}

\input{text/6_rq3_expressiveness_problems}

\subsubsection{Answer to RQ2}

For pure data retrieval, QPM matches and exceeds the expressiveness of basic query languages.
Our approach excels at expressing first-order logic conditions over graph structures of any complexity.
All query language features that are required for identifying quality problems are supported by QPM.
A comprehensive data quality analysis requires more than just the expressiveness of query languages; external access and natural language processing are also needed.
QPM surpasses this level of expressiveness by supporting custom Java functions, which enables these features.

%% file: text/6_rq3_expressiveness_languages.tex
\subsubsection{Expressiveness in Terms of Query Language Features (RQ2.a)}
\label{sec_evaluation_query_languages}

In answering RQ2.a, we compare the expressiveness of QPM with that of common database query languages.
We focus on XQuery, SPARQL, and Cypher because they are supported by our proof-of-concept implementation.
However, we also consider other query languages, such as SQL.

\paragraph{Language Features for Data Querying}

\autoref{subsec_querycomparison} discusses the features of query languages that are important for a comprehensive quality analysis.
\autoref{tab_querylanguagecomparison} shows an overview over the features across query languages.

The QPM language can express \emph{first-order logic} expressions using existential and universal quantifiers.
All \emph{traversal operators} are supported, including path expressions and property access.
These operators are essential for extracting information from a database.
QPM supports count operators that express cardinality constraints using subpatterns and numerical conditions.
The QPM language supports {\em logical operators and value comparators}.
Additionally, QPM supports type conversions to interpret values as different data types, such as numbers or timestamps, and compare them.
For strings, the \emph{Contains} operator checks whether a string contains a substring.
There is also a \emph{Match} operator that checks a string against a regular expression.
A major challenge is that some analyses require a semantic understanding of the data, which goes beyond the capabilities of traditional query languages (cf. \autoref{subsec_querycomparison}).
Our extension to custom operators (see \autoref{sec_implementation_java}) enables the integration of further analysis techniques, including Natural Language Processing (NLP) techniques as well as access to external databases and ontologies.

\paragraph{Further Operator Categories}

Apart from the features mentioned in \autoref{subsec_querycomparison}, query languages support additional operators that are not essential for data analysis.
To this end, we analyzed the language documentations of query languages\textsuperscript{\ref{xquery},\ref{sparql},\ref{cypher}} \cite{sql} and extracted some additional operator categories.
For this paper, we focus on data querying, despite the fact that all of these languages also include data transformation capabilities.

Currently, \emph{arithmetic operators} are not supported because they are unnecessary for identifying quality issues.
However, the QPM design allows for straightforward extension, which will enable future integration to support use cases such as metric calculations.

Query languages implement {\em function call operators} that can be both predefined methods or custom functions.
Since predefined methods are specific to query languages, QPM cannot support them universally.
Furthermore, not all common query languages support custom functions.
Nevertheless, we have developed a general solution for QPM that uses predicate filtering to extract parameters using query languages and execute such functions in a higher-level programming language (see \autoref{sec_implementation_java}).
This approach enables QPM to support a wide range of custom functions, including validating external links, DOIs, IP addresses, and spelling.
Our proof-of-concept implementation currently supports Java methods of the form $string \rightarrow boolean$, an important class of custom methods for value validation.
Future work includes adding more options.

The result of a query is either a list of values or a list of structured elements.
The QPM language supports aggregation and filtering operators for these lists.
However, {\em operators for grouping and ordering} are not yet supported, since there has been no identified need for grouping quality issues.
QPM queries are written so that they always require the complete set of findings.
The chronological default ordering is useful for the data improvement process and eliminates the need for ordering operators.
Once the analysis is finished, additional ordering and filtering can be handled at the application layer, where user-specific representations can be managed appropriately.

%% file: text/6_rq3_expressiveness_problems.tex
\subsubsection{Expressiveness with Respect to the Detection of Quality Problems (RQ2.b)}
\label{sec_evaluation_quality_problems}

In \autoref{sec_quality-problems}, we reported on our catalog of data quality problems in the cultural heritage domain in \cite{comprehensive_problem_specifications}.
From this catalog, we extracted categories of data quality problems, such as \emph{missing data}, \emph{heterogeneity of data}, and \emph{wrong data}.
In the following, we discuss how well QPM is able to detect these data quality problems.
\autoref{tab_comp_quality_problems} lists these data quality problem categories and maps them to the query language features needed to specify quality analyses for identifying such data quality problems.
\autoref{sec_evaluation_query_languages} shows that our approach is able to analyze each problem category from this list with the current implementation.

Navigating through the data and evaluating first-order logic is essential for identifying structural data patterns in the data.
For filtering data, our approach provides regular expressions and comparators.
QPM also supports counting the occurrences of elements and substructures, just as traditional query languages.

\input{text/6_evaluation_table-comp-quality-problems}

%% file: text/6_evaluation_table-comp-quality-problems.tex
\begin{table*}
	\centering
	\caption{Mapping of data quality problems from the catalog in \cite{comprehensive_problem_specifications} to query language features.}
	\label{tab_comp_quality_problems}
	\setlength{\tabcolsep}{2pt}

	\begin{tabular}{@{}lccccccl@{}}
	\toprule
	\textbf{Problem} & \textbf{FOL} & \textbf{Co} & \textbf{Cm} & \textbf{RE} & \textbf{EA} & \textbf{NLP} & \textbf{Catalog IDs from \cite{comprehensive_problem_specifications}}           \\ \midrule
	Missing Data     & \checkmark   & \checkmark  &             & \checkmark  & \checkmark  &              & D1, D4.2, D5, D6.7, D7.1, D7.2 \\
	Heterogeneity    & \checkmark   & \checkmark  & \checkmark  & \checkmark  &             &              & D1.3, D6.8                     \\
	Wrong data       & \checkmark   &             & \checkmark  & \checkmark  & \checkmark  & \checkmark   & D2, D7.3                       \\
	Misspelling      &              &             &             & \checkmark  & \checkmark  & \checkmark   & D2.1                           \\
	Inconsistency    & \checkmark   &             & \checkmark  & \checkmark  &             &              & D2.5, D4.1                     \\
	Dependency V.    & \checkmark   &             &             &             &             &              & D2.5, D5                       \\
	Redundancy       & \checkmark   &             & \checkmark  &             &             &              & D3                             \\
	Imprecision      &              & \checkmark  & \checkmark  & \checkmark  & \checkmark  &              & D6                             \\
	Format V.        & \checkmark   &             & \checkmark  & \checkmark  & \checkmark  &              & D2.3, D10, D11, D12            \\ \bottomrule
	\end{tabular}
	\\[0.2cm]
	
	\textbf{V.}: violation,
	\textbf{FOL}: first-order logic, 
	\textbf{Co}: count (metric),
	\textbf{Cm}: comparisons, 
	\textbf{RE}: regular expressions,
	\textbf{EA}: external access,
	\textbf{NLP}: natural language processing
	\\
	(EA and NLP are not supported by common query languages)
\end{table*}

%% file: text/6_rq4_useability.tex
\subsection{Usability (RQ3)}
\label{sec_evaluation_rq4}

To evaluate the usability of our approach, we conducted a qualitative usability study to explore the experiences of domain experts using our \lstinline|Constrainify| frontend.
Our study focused on the quality analysis of the supported LIDO XML format; therefore, we sought domain experts from the cultural heritage field who were familiar with the format.
Using a user-centric approach adopted from Lazar et al. \cite{lazar}, we tasked eight domain experts to define two specific new constraints for given data and perform a corresponding data quality analysis in scenario-based exercises \cite{zenodo-referencedata}.
The domain experts were data curators from different organizations with at least one year of experience handling LIDO data.

The domain experts had to define two constraints using \lstinline|Constrainify|:
One constraint was required to verify the existence of an identifier, and the other was required to verify the references of an image file.
We observed task interactions and results during the exercises and followed up with a structured questionnaire \cite{zenodo-questionaire} and open-ended interviews.
The questionnaire focused on the understandability and ease of use of Constrainify, while the interviews targeted usability and institutional fit.

\emph{The participants completed the workflows for both tasks and specified correct constraints.}
Observations and questionnaires revealed minor hesitations when selecting QPM templates for instantiation, searching for elements, and reviewing data excerpts of identified quality problems.
Usability emerged as a central theme, with participants emphasizing the modern UI and ease of use.
One participant noted that \q{quality reports are designed very intuitively and are clearly arranged}.
Rather than identifying barriers, the participants noticed two minor bugs in the UI that were resolved using our issue tracker.
They also suggested multiple features, including UI improvements and the addition of new QPM templates.
Key feature requests include providing hints on how to solve quality issues and advanced QPM templates using an if-then-else approach.
All suggestions have been prioritized for implementation in preparation for the productive operation of our tool.

The main {\em threats to the validity} of this study design~\cite{wohlin} concern construct, conclusion, and external validity.
Evaluating our own tool in a non-anonymous online setting may have introduced hypothesis guessing, evaluation apprehension, and experimenter expectancy effects.
The small sample size of eight participants reduces statistical power; however, this is less of an issue in qualitative studies.
The generalizability of the results is also restricted because the domain experts were selected as a homogeneous sample of only German organizations in the cultural heritage domain.
Future studies should include participants from different backgrounds and domains.
Additionally, a larger sample size would enable representative metrics.

\subsubsection{Answer to RQ3}
The user study indicates, that the tool effectively enables domain experts to define constraints and perform quality analyses.
Participants were able to complete the specified workflows according to the specifications and highlighted the modern interface.
Although there were some minor hesitations, no significant usability barriers were identified.
The feedback mainly consisted of suggestions for enhancements rather than issues.
Based on these findings, we cautiously conclude that the interface can effectively support domain experts in defining constraints and performing quality analyses.

%% file: text/7_related-work.tex
\section{Related Work}
\label{sec_related-work}

We aim to develop an approach that can enable domain experts to define domain-specific data quality analyses independently of specific database technologies.
This approach would eliminate the need for deep technical expertise and prevent the need for defining quality analyses several times for different database technologies.
Considering related work, this goal raises the following questions:
\begin{enumerate}[leftmargin]
	\item[\textbf{RQ4:}]
	What other generic query languages are available for querying databases? 
	\item[\textbf{RQ5:}]
	What other template- and pattern-based approaches exist for analyzing data quality?
\end{enumerate} 
In this section, we analyze the related work to answer these questions.

\subsection{Generic Query Languages}
\label{subsec_generic-querylanguages}

Generic query languages are applicable to more than one database technology. 
We found two generic query languages in the literature. 

Ong et al. \cite{sqlplusplus} note that data in NoSQL databases is typically modeled in JSON format.
Therefore, they developed SQL++ \cite{sqlplusplus}, which is a query language designed to work with schema-optional data stored in JSON format.
The syntax of SQL++ is based on SQL syntax, but it was extended to include the \lstinline|FROM|, \lstinline|WHERE|, \lstinline|GROUP BY| and \lstinline|SELECT| clauses, inspired by XQuery.
The semantics of SQL++ extend the expressiveness of SQL by removing semantic constraints regarding purely relational data structures.
In addition,``SQL++ semantics do not require schema or any homogeneity on the input data''~\cite{sqlplusplus}.

GraphQ is an approach that defines a ``unified intermediate representation for graph query languages''.
Nie et al. ~\cite{graphq} defined a new textual graph query language that incorporates modern English while maintaining fundamental graph structures.
A dedicated compiler can translate queries from this language into established graph query languages, such as SPARQL and Cypher.

\emph{In summary, although there are two query languages that show some degree of generality with respect to database technology, neither is truly generic.}
Both only support true subsets of common database technologies:
SQL++ uses the SQL syntax only to query JSON files, and GraphQ is limited to graph query languages.
In contrast, QPM is designed to support all common database technologies, focusing on XML, RDF, Neo4j, and RDB.
With these, QPM covers three types of database technologies: hierarchical, graph, and relational.

\subsection{Template-based Approaches}
\label{subsec_template-appraches}

\begin{figure}
	\centering
	\includegraphics[width=\linewidth]{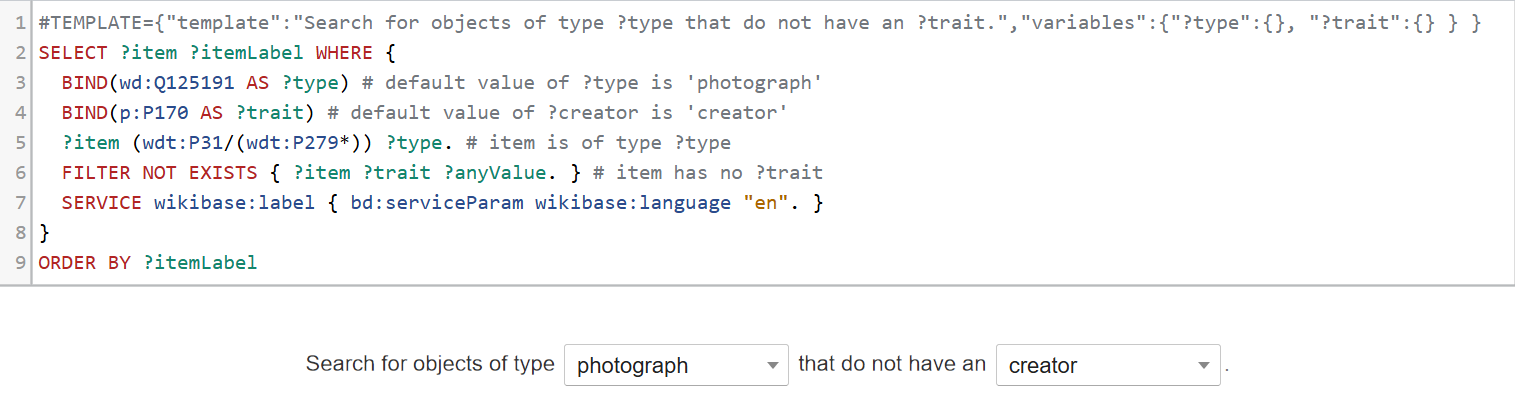}
	\caption{A Wikidata SPARQL template query to find photographs without creator, \url{https://w.wiki/GC4B}.}
	\label{fig_sparqltemplate}
\end{figure}

We have found the following template-based approach:
The Wikidata Query Service offers an approach to defining query templates using natural language for SPARQL templates\footnote{\url{https://en.wikibooks.org/wiki/SPARQL/Templates} (2026-06-25)}, which can be applied to Wikidata\footnote{\url{https://query.wikidata.org} (2026-06-25)}.
A SPARQL template is defined by providing a query and a natural language phrase containing the same variables.
An example regarding paintings without an artist (creator) is presented in \autoref{fig_sparqltemplate}.
It starts by defining the template text using variables (here $?type$ and $?trait$).
The following query defines the search logic.
The Wikidata user interface provides a query helper that renders the template sentence with a drop-down menu, as shown below the query.
It is possible to narrow down the options in the drop-down menu by providing additional queries per variable.
However, this approach is limited to the Wikidata RDF database and can exclusively be used via their website.
Furthermore, the syntax of the templates is very restricted, which leads to inefficient querying.
This demonstrates the viability and practicality of template-based query specification, but only for very limited use cases.
We identified no approaches in the scientific literature regarding natural language templates for query specification.

In summary, QPM extends the template concept of the Wikidata user interface by offering support for flexible templates, multiple databases, and database technologies.

\subsection{Pattern-Based Approaches for Data Quality Analysis}
\label{subsec_related-work_pattern-based}

We have found several approaches that use pattern-based analysis to detect data quality problems.
In general, patterns define the structures to be searched for and serve the same function as templates in identifying constraint violations.
For each pattern-based approach, we determine the level of abstraction and measure the expressiveness using the features discussed in \autoref{sec_evaluation_quality_problems}.
We also consider the language used to define the patterns.
Regarding the levels of abstraction, we consider the three levels:
\q{Concrete} approaches define specific queries,
\q{abstract} approaches define query templates specific to a database technology, and 
\q{generic} approaches that additionally abstract from the technologies result in technology-independent query templates.
An overview is given in \autoref{tab_relatedWork}.
The approaches are discussed below. 

Both Kontokostas et al. \cite{kontokostas2014test} and Fürber et al. \cite{furber2010using, furber_swiqa_2011} presented similar approaches to detect quality problems in Linked Data.
These approaches are based on parameterized query templates for SPARQL and SPIN \cite{spin}.
Each of these pattern-based approaches is designed for a specific database technology.
These approaches have similar potential and limitations to ours.
However, both approaches require abstract patterns (i.e. queries) to be written manually in SPARQL or SPIN.
Thus, the queries contain more implementation details than queries in QPM.

Bizer et al. \cite{bizer_quality-driven_2009} presented a policy framework for quality-driven information filtering in graph-based data.
They only consider patterns at the concrete level.
The approach is limited to propositional logic and does not support counting pattern occurrences in the data.
Patterns are expressed using a custom SPARQL-based language.

Bicevska et al. \cite{oditis_domain-specific_2017} proposed using a domain-specific language (DSL) to specify data quality requirements for specific database technologies and formats.
Thus, only concrete patterns are considered.
This approach differs slightly from the above in terms of the features supported.
The pattern language, i.e., the DSL, is not presented in the paper.
Instead, informal explanations are used to describe examples for quality specifications.
The authors proposed to translate them into a query language, such as SQL, but did not present an algorithm. 
To motivate their approach, Bicevska et al. discussed the use of the object constraint language (OCL) to define data quality.
OCL is powerful enough to specify constraints in first-order logic and beyond.
However, it is not well-suited for domain experts without good skills in object-oriented programming.
In addition, OCL is fully typed, making it well-suited for structured data; however, research data is often semi-structured~\cite{semistructured_data}.

\input{text/7_table_pattern-approaches.tex}

When comparing the expressiveness of these approaches to the query languages (cf. \autoref{subsec_querycomparison}), we find that the approaches do not surpass the features.
In fact, some approaches even have limited expressiveness compared to the base query language.
They do not offer custom operators, and especially not access to external data or NLP techniques.
Furthermore, all approaches are limited to one specific database technology.

\emph{In summary, QPM is the only approach that supports the definition of data quality analysis at a generic level}, making it highly flexible with respect to underlying database technologies.
The proof-of-concept implementation of QPM already demonstrates the applicability to XML, RDF and Neo4j databases.
This establishes the basis for a user-friendly representation of quality analysis at different levels of abstraction.
Additionally, \emph{QPM incorporates access to external data and NLP techniques} (see \autoref{sec_implementation_java}).
In this way, QPM exceeds the expressiveness of all pattern-based approaches and query languages.

\subsection{Model-Driven Approaches to Data Quality Analysis}
\label{subsec_related-work_mde}

There are a few approaches to data quality analysis using model-driven engineering (MDE).
In \autoref{subsec_related-work_pattern-based}, we discussed the approach by Bicevska et al. \cite{oditis_domain-specific_2017}, which uses a DSL to specify data quality requirements.

Nikiforova et al. \cite{mde_nikiforova} also propose an MDE approach for specifying data quality requirements.
They present a two-step model, in which the requirements are first specified in terms of an informal platform-independent model (PIM) using natural language and a graphical flowchart-based diagram.
Then, a platform-specific model (PSM) is created by manually translating these requirements into executable artifacts.
Compared to our approach, theirs does not require any formalization at the platform-independent level, but the instantiation must be done manually.

Karkouch et al. \cite{mde_karkouch} propose an MDE approach to generate a customized infrastructure for data quality management.
Using an Eclipse-based model editor, domain experts can specify data quality requirements.
From the defined requirements, a complete database management system is automatically generated that ensures ``the capture, computing and persisting or/and streaming of [data quality] information'' \cite{mde_karkouch}.
The implementation includes the generation of a relational database in Oracle and a JavaSE application for monitoring data quality.
However, this work is not suitable for our scenario because it requires an existing application before data can be collected.
It is also limited to certain Oracle databases and cannot be applied retrospectively.

Multi-level modeling (MLM) \cite{totem, mlm_datainteroperability, mlm_extending_deep_meta_modelling, mlm_when_and_how_to_use, mlm_formalisation} also seems promising in the context of data quality analysis, as it can correctly represent the different levels of abstraction of database technologies and schemas.
{\em To the best of our knowledge, ours is the only approach that strictly models different abstraction levels using MLM for template-based quality assessment.}

\subsection{Summary}

To compare QPM with related work, we examined several lines of research.
Now, we can answer the questions we raised at the beginning of this section.

\subsubsection{Answer to RQ4}
There are approaches to extending the expressiveness of specific database query languages to additional database technologies.
However, we found no other query language as generic as QPM in terms of underlying database technologies and data models.
Additionally, QPM's extension to custom operators significantly increases its expressiveness compared to the approaches we considered.

\subsubsection{Answer to RQ5}
Regarding template-based approaches, we only found SPARQL templates for Wikidata, which are used to specify data queries.
However, these templates are limited in functionality.
There are also related pattern-based approaches; however, none of them are independent of database technology, nor do they provide effective means for domain experts to define quality analyses.
These are two key problems that QPM aims to solve.

%% file: text/7_table_pattern-approaches.tex
\begin{table}
	\setlength{\tabcolsep}{2pt}
	\centering
	\caption{Comparison of pattern-based approaches with query languages (c.f. \autoref{subsec_querycomparison}) and QPM}
	\begin{tabular}{p{0.22\textwidth}p{0.195\textwidth}%
			ccccccp{0.2\textwidth} }
		\toprule
		\textbf{Approach}                                                     & \textbf{Abstraction}    & \textbf{FOL} & \textbf{Co} & \textbf{Cm} & \textbf{RE} & \textbf{EA} & \textbf{NLP} & \textbf{Language}  \\
		\midrule
		& concrete & \checkmark   & \checkmark  & \checkmark  & \checkmark  &             &              &  \makecell[l]{\tiny SQL, XQuery, SPARQL, \\ \tiny Cypher (cf. \autoref{tab_querylanguagecomparison})}           \\
		\arrayrulecolor{gray!30} \midrule \arrayrulecolor{black}
		Kontokostas et al. \cite{kontokostas2014test}                         & abstract (RDF)          & \checkmark   & \checkmark  & \checkmark  & \checkmark  &             &              & SPARQL             \\
		Fürber et al. \cite{furber2010using, furber_swiqa_2011}               & abstract (RDF)          & \checkmark   & \checkmark  & \checkmark  & \checkmark  &             &              & SPIN               \\
		Bizer et al. \cite{bizer_quality-driven_2009}                         & \makecell[l]{concrete \\ (Named Graphs)} & PL           &             &             & \checkmark  &             &              & WIQA-PL            \\
		Bicevska et al. \cite{oditis_domain-specific_2017}                    & \makecell[l]{concrete \\ (relational)}   & \checkmark   &             & \checkmark  &             &             &              & graphical DSL, SQL \\
		\arrayrulecolor{gray!30} \midrule \arrayrulecolor{black}
		\textbf{QPM} & generic                 & \checkmark   & \checkmark  & \checkmark  & \checkmark  & \checkmark  & \checkmark   & DSL                \\ \bottomrule
	\end{tabular}
	\vspace{1mm}
	\caption*{
		\normalfont
		\textbf{FOL}: first-order logic,
		\textbf{PL}: propositional logic ($<$ FOL),
		\textbf{Co}: count (metric),
		\textbf{Cm}: comparisons, 
		\textbf{RE}: regular expressions,
		\textbf{EA}: external access,
		\textbf{NLP}: natural language processing,
		\textbf{DSL}: domain-specific language
	}
	\label{tab_relatedWork}
\end{table}

%% file: text/8_conclusion.tex
\section{Conclusion}
\label{sec_conclusion}

Quality analysis is central to data management, since ensuring the quality of data is essential for its effective use.
There is no general definition of data quality, as it is typically domain-specific.
Domain experts are primarily responsible for defining quality requirements for their type of data.
However, domain experts often lack the skills to realize domain-specific quality analyses independently and must therefore coordinate with data engineers.

To empower domain experts, we present \lstinline|QPM|, a model-driven approach to defining domain-specific data quality constraints independent of specific database technologies.
This approach can eliminate the need for deep technical expertise and prevent the need for defining quality analyses several times for different database technologies.
The QPM framework supports various data models and even several database technologies for data analysis.
Domain experts can specify their own data quality analyses and also update them as needed.
They can choose from a set of predefined QPM templates to define new domain-specific quality analyses.
QPM assists domain experts in instantiating selected QPM templates to meet their domain-specific needs, so only limited technical expertise is required.
This model-driven approach is realized as a proof-of-concept tool with the \lstinline|QPM| backend that supports the database technologies XML, RDF, and Neo4j.
To define and perform domain-specific data quality analyses, we added a frontend called \lstinline|Constrainify|.
\lstinline|Constrainify| currently works with XML data and provides specific support for the LIDO and TEI data formats.

We evaluated our approach in terms of applicability, expressiveness, and user-friendliness by defining domain-specific data quality analyses in the cultural heritage domain and applying them to real-world databases with up to 120 million data records.
Regarding data quality analysis, our approach matches the expressiveness of common data query languages, such as XQuery, SPARQL, and Cypher.
As an extension, QPM provides the facility to define custom functions using an external programming language, such as Java, for specifying advanced data quality analyses.
These functions allow for data analyses that are not native to query languages.
Examples include URL validation, natural language processing techniques, and access to external databases and ontologies.
We evaluated the usability of our tool with domain experts, who reported that it is user-friendly and allows them to specify comprehensive quality analyses on their data.

The framework is currently being integrated into real-world data processing workflows.
We are collaborating with domain experts from the Head Office of the GBV Common Library Network\footnote{\url{https://en.gbv.de/} (2026-06-25)} (VZG) and the German Digital Library\footnote{\url{https://www.deutsche-digitale-bibliothek.de} (2026-06-25)}.
Before integrating new data into these platforms, our tool ensures the quality of the data.
We help these institutions analyze and improve their data by providing them with up-to-date, domain-specific quality assessment techniques and tools.
We are extending our investigation to promising applications in other well-established collaborative platforms, with a current focus on cultural heritage domains, such as Europeana\footnote{\url{https://www.europeana.eu} (2026-06-25)} and the National Research Data Initiative consortia\footnote{\url{https://www.nfdi.de/consortia/} (2026-06-25)}.

In the future, we plan to expand the implementation, particularly the frontend, to support additional data formats and technologies.
We will extend \lstinline|QPM| to also support relational database technologies and JSON.
Furthermore, we plan to expand to other types of data beyond cultural heritage.
Thus, we aim to support other formats as well, such as the ABCD schema \cite{abcdschema} from the biodiversity domain.

%% file: text/0_acknowledgement.tex
\section*{Acknowledgement}

This paper was written as part of two projects: KONDA\footnote{``Kontinuierliches Qualitäts\-management von dynamischen Forschungsdaten zu Objekten der materiellen Kultur unter Nutzung des LIDO Standards'' (KONDA, 2019-2023) \url{https://zenodo.org/communities/konda-project/about}}
and AQinDa\footnote{``Agile Qualitätssicherung von Metadaten zu kulturellen Objekten im Kontext von Datenintegrationsprozessen'' (AQinDa, 2023-2027) \url{zenodo.org/communities/aqinda/about}}.
KONDA was funded by the German Federal Ministry of Education and Research (BMBF).
AQinDa is funded by the German Research Foundation (DFG).
The goal of these projects is to develop a continuous management process for improving data quality, with a focus on cultural heritage data.